\documentclass[aps,prc,showpacs,twocolumn,groupedaddress]{revtex4}

\usepackage{graphicx}
\usepackage{amsmath}
\usepackage{amssymb}
\usepackage{amsfonts}
\usepackage{color}
\usepackage{psfrag}
\usepackage{hyperref}
\usepackage{bm}
\graphicspath{{./img/}}

\newcommand\bn{\ensuremath{\bm{\nabla}}}
\begin{document}

\title{JM-ECS: A hybrid method combining the $J$-matrix and ECS methods for scattering calculations}

\author{Y.\ Bidasyuk}
\email[]{Yuriy.Bidasyuk@ua.ac.be}
\affiliation{Departement Wiskunde-Informatica, Universiteit Antwerpen, Antwerpen, Belgium}
\affiliation{Bogolyubov Institute for Theoretical Physics, Kyiv, Ukraine}
\author{W.\ Vanroose}
\email[]{Wim.Vanroose@ua.ac.be}
\author{J.\ Broeckhove}
\email[]{Jan.Broeckhove@ua.ac.be}
\author{F. Arickx}
\email[]{Frans.Arickx@ua.ac.be}
\affiliation{Departement Wiskunde-Informatica, Universiteit Antwerpen, Antwerpen, Belgium}
\author{V.\ Vasilevsky}
\affiliation{Bogolyubov Institute for Theoretical Physics, Kyiv, Ukraine}

\date{\today}

\begin{abstract}
The paper proposes a hybrid method for calculating scattering
processes. It combines the $J$-matrix method with exterior complex
scaling and an absorbing boundary condition. The wave function is
represented as a finite sum of oscillator eigenstates in the inner region
and it is discretized on a grid in the outer region. The method is validated
for a one- and two-dimensional model with partial wave equations, and
a calculation of $p$-shell nuclear scattering with semi-realistic interactions.
\end{abstract}


\maketitle

\section{Introduction  \label{intro}}
Numerical simulations of breakup reactions are among the most
complicated problems in the theoretical physics of nuclear
reactions. The main complication arises from the necessity of modeling
the asymptotic behavior of the wave function in a many-body continuum. To
simplify the asymptotic behavior, the wave function is usually
expressed in terms of hyperspherical harmonics
\cite{esry1996adiabatic,cobis1997computations,broeckhove20075h}. But,
in breakup problems, the hyperpotentials decay very slowly and
convergence with respect to the hyper-angular momentum is also very slow. This
results in very large systems of equations, even for the simplest breakup
problems.

The Complex Scaling approach
\cite{junker1982recent,moiseyev1998quantum} can be used to
avoid the problems associated with the asymptotic behavior of the many-body wave
function. However,  this method is suitable only to analyze resonances in
the system and it does not provide any breakup information, such as
cross sections, that can be compared with experiments.

In the numerical solution of breakup problems described by the
Schr\"odinger equation it is important to describe each of the breakup
channels in a correct way.  Once the asymptotic form in each of the
channels is understood, it is possible to formulate a set of equations
whose solution yields both the wave function in the interaction region, and
the asymptotic amplitudes in each breakup channel.
This is the approach taken both by the $R$-matrix \cite{lane1958r} and the
$J$-matrix method \cite{heller}.

However, for some of the breakup channels the asymptotic wave function
is only known at very large distance from the target. As a consequence
very large numerical domains are needed to cover the interaction
and the near-field regions.  In other cases, such as the three body
breakup in the $J$-matrix representation, the explicit expression of the
asymptotic wave function is not known. The need for an explicit asymptotic
wave fuction can be avoided by the introduction of absorbing boundary
conditions. These enforce an ``outgoing wave'' boundary condition on the
numerical solution, and have been used successfully in
atomic and molecular physics \cite{mccurdy2004} and acoustic and
electromagnetic scattering problems \cite{berenger1994perfectly}.
Instead of solving in a single system both the wave function in the
interaction region and the reaction rates, these
methods calculate the cross section as a post-processing step. First
the equations are solved with the absorbing boundary conditions and
then, in a second step, the amplitudes are extracted from the
numerical solution.

Absorbing boundary conditions are easy to implement in a grid based
discretization of a partial differential equation.  They have been
implemented in calculations based on finite difference, B-splines, finite elements etc
\cite{mccurdy2004, givoli}. 

In nuclear physics, however, one prefers to use a $L^2$-basis, such as
the oscillator eigenstates. Such basis is well-adapted to the fully
microscopic description of the compound nucleus. This paper reports on
the initial efforts to introduce this approach in the context of
nuclear few-body systems. We propose a hybrid approach with an $L^2$
representation of the wave function in the interaction region, and
with a discretized grid representation in the outer region. Our aim is
to combine the strengths and benefits of both approaches.

The paper is structured as follows. In section II, after a short review
of the ECS and $J$-matrix method, we introduce the hybrid method where
the wave function is represented in the inner region by oscillator
states and in the outer region by a grid. We also discuss how the
observables are extracted from the numerical representation of the
scattered wave. In section III we present numerical results that validate
the proposed method with model problems for one- and two-dimensional
partial wave equations. In section IV we show results for
nuclear $p$-shell scattering.  We use $m=1$ and $\hbar=1$ throughout
the paper, except where mentioned.


\section{The hybrid $J$-matrix with ECS method}

\subsection{Exterior Complex Scaling as an outgoing wave boundary condition}
\label{sec:ecs}
\textit{Exterior Complex Scaling} (ECS) was introduced by B. Simon
\cite{simon1979definition} and has been initially used for the
calculation of resonance positions and widths
\cite{moiseyev1998quantum}. It has also found widespread application
by providing an absorbing boundary condition in atomic and molecular
breakup problems with charged particles
\cite{rescigno1999collisional,vanroose2005complete}.  A review is
given in \cite{mccurdy2004}.  In these problems the outgoing wave is
very complicated, and its form depends on the effective interaction of
the outgoing particle. It is determined by the position of the other
charged particles involved in the breakup problem.

The derivation of ECS 
is based on an analytical
continuation in the complex plane.  It differs from the well-known
\textit{Complex Scaling} (CS) by starting the scaling procedure well
into the asymptotic region.

We briefly explain why it leads to outgoing wave boundary conditions
and apply these to the Schr\"odinger equation.

\subsubsection{Enforcing outgoing wave boundary conditions}\label{sec:enforcingdirichlet}
Consider a one-dimensional Helmholtz equation
\begin{equation}
  \left(-\frac{d^2}{d\rho^2} - k^2\right) u(\rho) =
  f(\rho) \label{eq:helmholtz}
\end{equation}
on a domain $[0,L]$, with constant wave number $k > 0 \in \mathbb{R}$,
homogeneous Dirichlet conditions on the left boundary, and outgoing
wave boundary conditions on the right boundary.  The right hand side,
$f(\rho)$, is such that it is zero outside the interval $[0,a]$ with
$a < L$.

Let us focus this discussion on the right boundary in $L$. For $a <
\rho < L$ the equation is a second order homogeneous differential
equation with constant coefficients.  In this region the general
solution can then be written as
\begin{equation}
  u(\rho) = A e^{ik\rho}  + B e^{-ik\rho},  \label{eq:fundamental solutions}
\end{equation}
a linear combination of two fundamental solutions.  Enforcing outgoing
wave boundary conditions in $\rho=L$ means that coefficient $B$ should
be zero.  This can be realized by the following mixed type boundary condition on the
solution
\begin{equation}
  u^\prime(L) = i\, k \, u(L). \label{eq:mixed_boundary_condition}
\end{equation}
Note that this boundary condition requires the explicit knowledge of
the wave number $k$.

ECS is an alternative way to enforce the same outgoing wave condition.
When we analytically continue the equation \eqref{eq:helmholtz} to
complex $\rho\in \mathbb{C}$,  the general solution of the equation,
where it is homogeneous, remains a linear combination of the same
fundamental modes as in \eqref{eq:fundamental solutions}.  When we
impose, instead of the condition \eqref{eq:mixed_boundary_condition},
a homogeneous Dirichlet boundary condition in the point $L^\prime \in
\mathbb{C}$ that lies inside the region where the equation is
homogeneous, we find that $u(L^\prime)=0$ leads to $B/A = \exp(2ik
L^\prime)$ or
\begin{equation}\label{eq:incoming_suppression}
  \frac{|B|}{|A|} = \exp(-2\, k \,\mbox{Im}(L^\prime)).
\end{equation}
So if the point $L^\prime \in \mathbb{C}$ is chosen such that
Im$(L^\prime) > 0 $ and $k\,$Im$(L^\prime) \gg 1$ then $|B| \ll
|A|$. As a result, we have, effectively, enforced outgoing wave
boundary conditions in the point $L^\prime$.

The linear combination of fundamental modes with coefficients $A$ and
$B$ describes the solution everywhere where the equation
\eqref{eq:helmholtz} is homogeneous. The equation is homogeneous
between the point $L$ and $L^\prime$.  So the coefficient $B$ of the
solution is still much smaller than $A$ in the point $L$ and we can
conclude that we also have an outgoing wave boundary condition in $L$.

It is important to note that, in contrast to the mixed boundary
condition \eqref{eq:mixed_boundary_condition}, enforcing Dirichlet
boundary conditions in $L^\prime$ does not require the knowledge of
$k$. This makes it possible to describe both inelastic processes and
outgoing boundary conditions in higher dimensions.

\subsubsection{ECS contour}
We implement this by extending the interval $[0,L]$ with a contour
that connects $L$ with $L^\prime$.  The point $L^\prime$ is typically
chosen such that $|L| < |L^\prime|$ and $k |L^\prime| \gg 1$.

Such contour can be formulated as a coordinate transformation on the
interval $[0,\,$Re$(L^\prime)]$.
\begin{equation}
 z(\rho) = \left\{ \begin{aligned}
		&\rho   &\quad&\text{if} \quad \rho \le L.\\
		&\rho + i f(\rho)&\quad&\text{if} \quad L < \rho <  \text{Re}(L^\prime).
             \end{aligned}\right.
\end{equation}
where $f$ is an increasing function --- linear, quadratic, \ldots ---
with $f(Re(L^\prime))=Im(L^\prime)$ and $f(L)=0$.

In particular, ECS considers a linear function, and the coordinate
transformation is therefore written as
\begin{equation}
 z(\rho)  = \left\{
 \begin{aligned}
 &\rho &\quad& \text{for}\quad \rho\le L. \\
 &L  + (\rho-L) \mathrm{e}^{\mathrm{i} \theta} &\quad& \text{for}\quad \rho>L.
 \end{aligned}
 \right. \label{eq:ecstrafo}
\end{equation}

\subsubsection{Application to the Schr\"odinger equation}
The application of these outgoing wave boundary conditions to the
Schr\"odinger equation is straightforward. Consider, as an example, the
two-particle model problem for which the radial equation is
\begin{equation}\label{eq:onedimensional}
  (T + V(\rho) + \frac{l(l+1)}{2\rho^2}-E)\psi_{sc}(\rho) = - V \hat{j}_{l}(k \rho), 
\end{equation}
where $T=(-1/2)\,d^2/d\rho^2$ denotes the kinetic energy operator,
$k=\sqrt{2E}$ is the wave number, $l=0,1,2,\ldots$ the angular
momentum, $V$ the potential that only depends on the radial coordinate
$\rho$, and $\hat j_{l}(k\rho)$ the Ricatti-Bessel function, which is
the radial part of the incoming wave. The radial solution of this
equation is $\psi(\rho) = \hat{j}_l(\rho) + \psi_{sc}(\rho)$, a sum of
the incoming wave, $\hat{j}_{l}(\rho)$, and the scattered wave,
$\psi_{sc}(\rho)$. The latter should fit the outgoing wave boundary
conditions.

For problems with short range potentials the Schr\"odinger equation
reduces to the Helmholtz equation outside the range of the
potentials with a wave number $k$.  We can then apply the
ECS transformation in the region where it reduces to a Helmholtz equation.

\subsubsection{Discretization}
\begin{figure}[ht]
\includegraphics[width=0.95\linewidth]{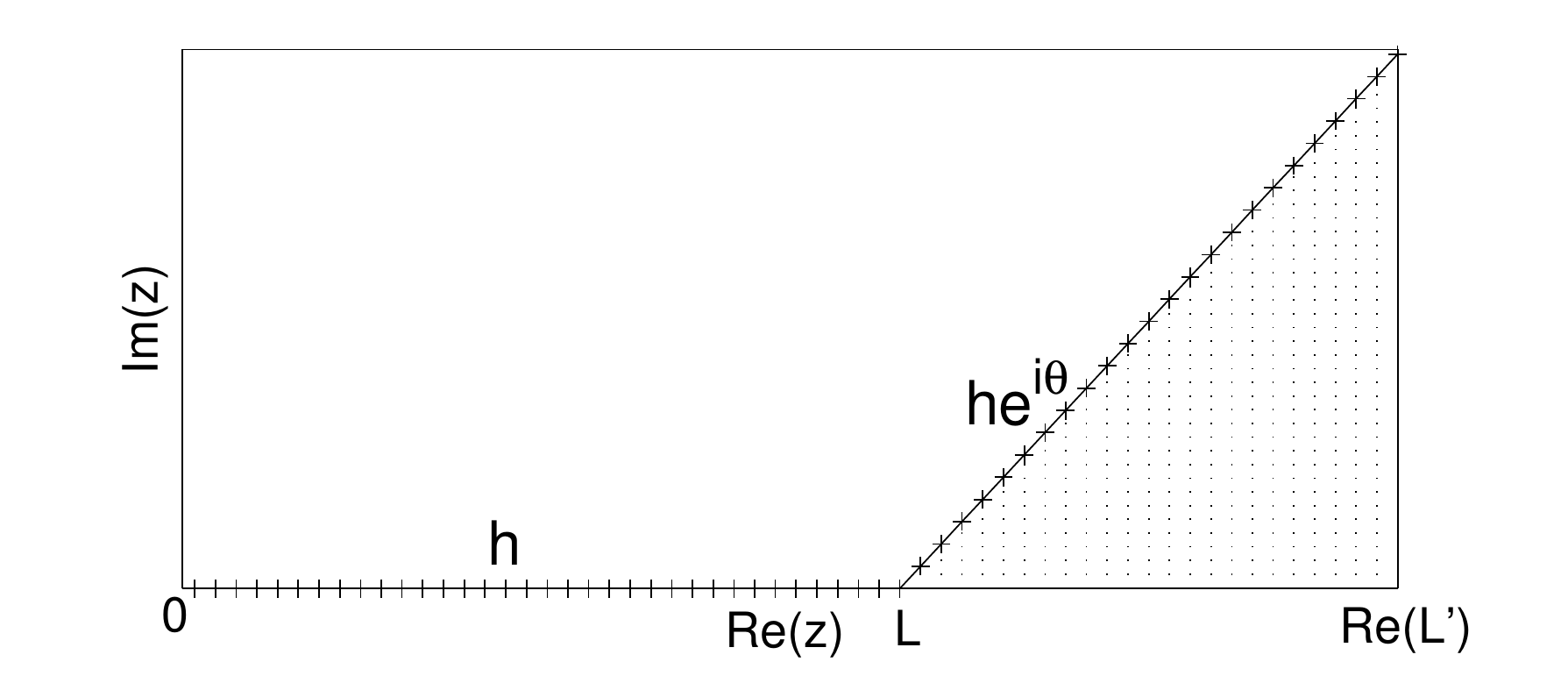}
 \caption{The choice of grid points for the finite difference
   representation of the Helmholtz equation on a exterior complex
   scaled (ECS) domain. The original problem was stated on $[0,L]$ with
   outgoing wave boundary conditions in $L$. The domain is extended
   with $[L,L^\prime]$ where a complex grid distance $h\,e^{i\theta}$ is used. In $L^\prime$ homogeneous Dirichlet boundary
   conditions are enforced.
\label{fig:contour}}
\end{figure}

ECS has been implemented in finite differences, B-splines and spectral
elements \cite{mccurdy2004}.  In this section we discretize the
differential operator that appears in the Schr\"odinger equation with
finite differences using the Shortley-Weller formula for non-uniform
grids \cite{shortley}. It enables the discretization of the operator on the
complex contour and uses complex valued mesh widths. Such a mesh mesh
is illustrated on Fig.~\ref{fig:contour}

Consider the differential operator $d^2/d{z(\rho)}^2$ on the contour
defined by the coordinate transformation \eqref{eq:ecstrafo}.  We
define a uniform grid
\begin{equation*}
(z_i)_{0\leq i\leq n} \mbox{ on } [0,L]
\end{equation*}
with $z_0 = 0$ and $z_n=L$ and mesh width $h = 1/n \in\mathbb{R}$, and a second uniform grid
on the complex contour
\begin{equation*}
(z_i)_{n\leq i\leq n+m} \mbox{ on } [L,L^\prime]
\end{equation*}
with $z_{n+m} = L^\prime $ and complex mesh width $h_\gamma = (L^\prime-L)/m$. The union of these two grids is the grid
\begin{equation}\label{eq:ecsgrid}
(z_i)_{0\leq i\leq n+m} \mbox{ on } [0,L]\cup[L,L^\prime]
\end{equation}
in the entire ECS domain.  To approximate the second derivative in
each grid point $z_i$ we use
\begin{equation}\label{eq:shortwell}
\begin{split}
&\frac{d^2u(z_i)}{d z^2} \approx
\frac{2}{h_{i-1}+h_i} \\
&\times \left(\frac{1}{h_{i-1}}u(z_{i-1})-\left(\frac{1}{h_{i-1}}+\frac{1}{h_i}\right)u(z_i)
+\frac{1}{h_i}u(z_{i+1})\right),
\end{split}
\end{equation}
where $h_{i-1}$ and $h_i$ are the left and right mesh widths
respectively, which may be complex valued.  The formula reduces to
regular second order central differences when $h_{i-1}=h_i$, i.e., in
the interior real region $[0,L]$, and in the interior of the complex
contour $[L,L^\prime]$, since the scaling function $f$ is taken to be
linear.  The only exception is the point $z_n$ where we lose at most
an order of accuracy.  However, with ample discretization steps, the
overall accuracy is anticipated to match up to second order. Note that
a higher order discretization at the hinge is proposed in
\cite{baertschy2001electron} to maintain a second order accuracy
throughout the domain.

The Hamiltonian of the one-dimensional model \eqref{eq:onedimensional}
is then a $(n+m)$ by $(n+m)$ matrix $\mathsf{H}_l \in
\mathbb{C}^{n+m}$.
\begin{equation}
  \mathsf{H}_l = \mathsf{T} + \frac{l(l+1)}{2}\texttt{diag}(1/z_i^2) +
  \texttt{diag}(V(z_i)),
\end{equation}
where $\mathsf{T}$ is the finite difference representation of the
second derivative on the complex contour, and \texttt{diag} is a
diagonal matrix with the potential evaluated in each grid point $z_i$.

For two-dimensional problems, the Hamiltonian is discretized starting from 
the one-dimensional discretized Hamiltonian using the Kronecker
product $\otimes$
\begin{equation} \label{eq:twoDECS}
  \mathsf{H}_{2D} =  \mathsf{H}_{l_1} \otimes \mathsf{I} + \mathsf{I} \otimes  \mathsf{H}_{l_2} +  \texttt{diag}({V}_{12}(z_i,z_j)),
\end{equation}
where $\mathsf{I}$ is a $(n+m)$ by $(n+m)$ unit matrix, and
\texttt{diag}$(V)$ is the two body potential evaluated at the grid
points.  The matrix $\mathsf{H}_{2D}$ is now a complex valued
$(n+m)^2$ by $(n+m)^2$ matrix.

The numerical solution of the PDE \eqref{eq:onedimensional} is found by
solving the linear system
\begin{equation}
  (\mathsf{H}_l-E)\mathsf{x} = \mathsf{b},
\end{equation}
where $\mathsf{b}$ is a numerical representation of the right hand
side of the equation.

\begin{figure}[ht]
\includegraphics[width=0.9\linewidth]{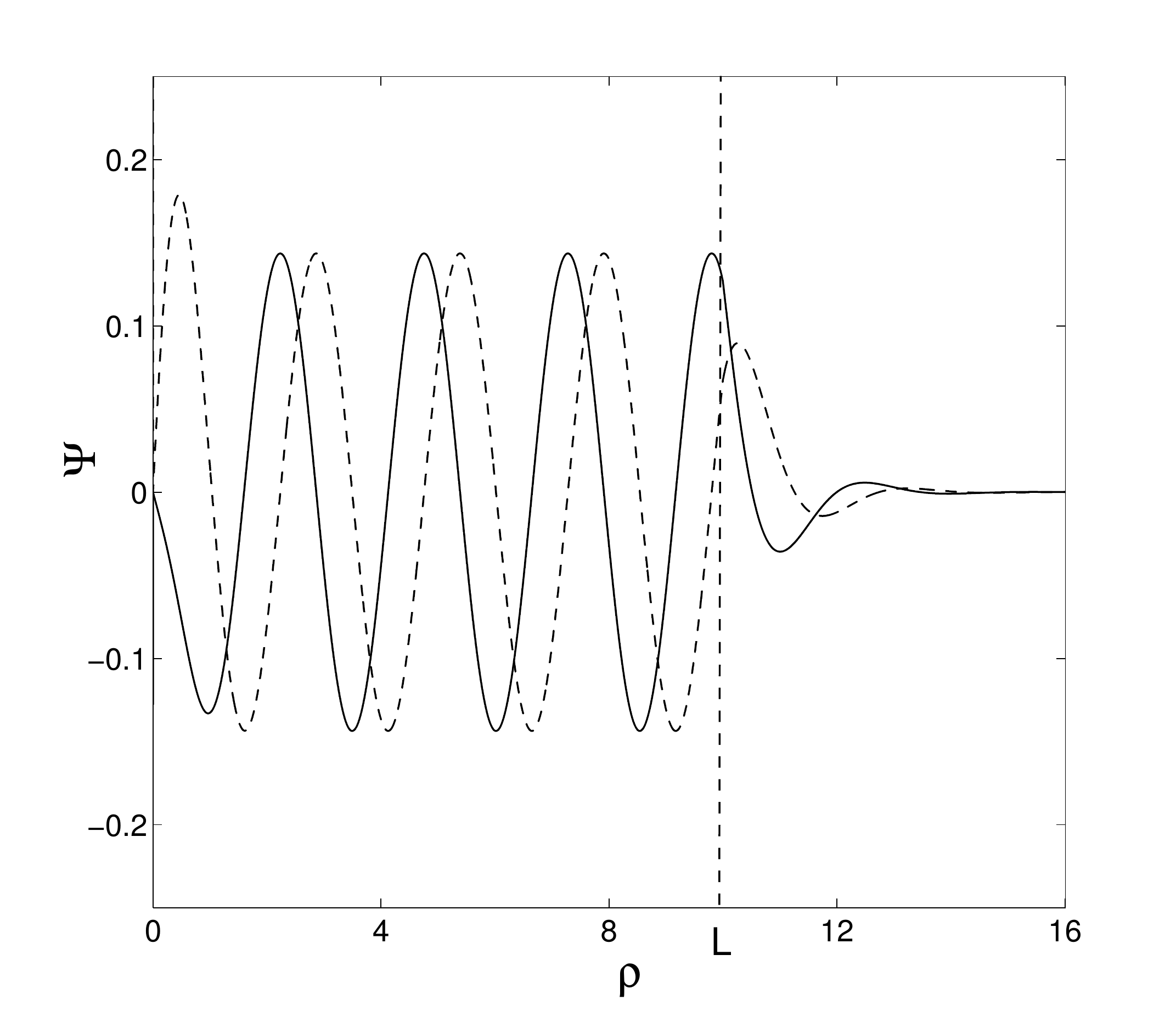}
 \caption{ The real and imaginary part of the scattered wave for an
   Helmholtz problem \eqref{eq:helmholtz} on the ECS grid. Because of
   the homogeneous Dirichlet boundary condition the wave is only
   outgoing in $L$.}
\label{fig:sample_1d}
\end{figure}

\begin{figure}[ht]
\includegraphics[width=0.9\linewidth]{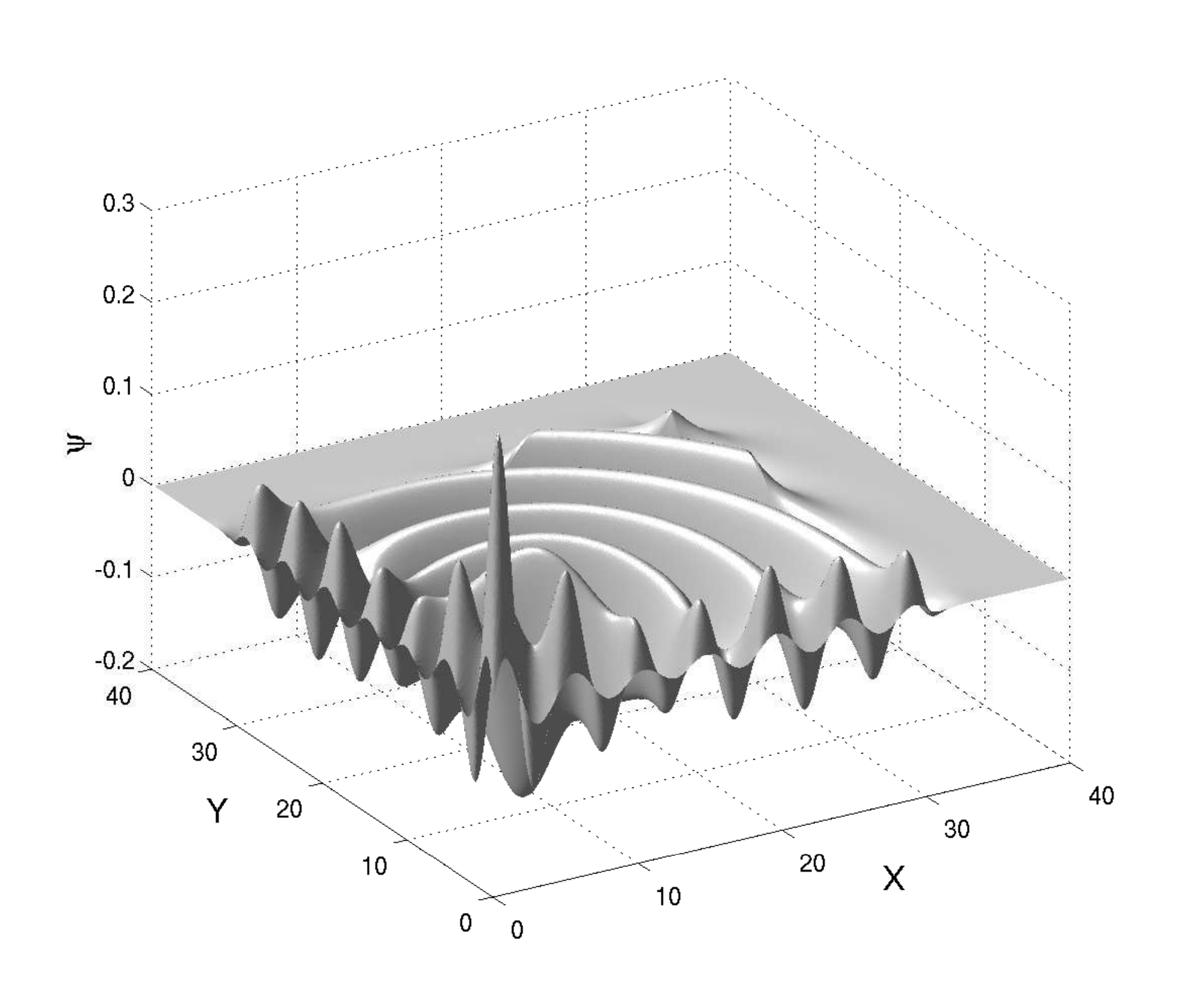}
\caption{The real part of a 2D partial wave of a three-particle
  $s$-wave scattering problem on an ECS domain, \eqref{eq:2d_1}. The
  problem has a homogeneous Dirichlet boundary condition on the domain
  $[0,L^\prime]^2$ and because of the ECS this leads to a problem on
  the domain $[0,L]^2$ with homogeneous Dirichlet boundary condition on the south and west
  boundary, and outgoing wave boundary conditions on the north and east
  boundary where $x=L$ or $y=L$. We show the numerical solution of the
  problem discretized with finite differences.}
\label{fig:sample_2d}
\end{figure}

Some examples of one- and two-dimensional wave functions on the ECS
domain are presented in Fig.~\ref{fig:sample_1d} and
\ref{fig:sample_2d}.


\subsubsection{Extraction of the observables}\label{sec:extract}
Once we have the numerical solution of the equation on an ECS domain,
we need to extract physical observables such as cross sections.

In one dimension the amplitude $A$ of the outgoing wave can be
extracted from the numerical solution with the help of the
Wronskian. Indeed, outside the range of the potential $V$ the solution
can be written as a linear combination $\psi_{sc} = A
\hat{h}_l^+(kr) + B \hat{h}_l^-(kr)$, where $\hat{h}_l^\pm$ are the
in- and outgoing Ricatti-Hankel functions. The coefficient $A$ is then
\begin{equation}
 A = W\left(\psi_\text{sc}(\rho),\hat{h}_l^-(k\rho)\right)/W\left(\hat{h}_l^+(k\rho),\hat{h}_l^-(k\rho)\right),
\end{equation}
where $\rho \in [a,L]$ is outside the range of the potential but still
on the real part of the ECS domain. The Wronskian is calculated as
$W(u,v) = u^\prime v-v^\prime u$. Since only the values of $\psi_{sc}$
are known on the grid points, its first derivative is approximated
by central differences.

In two dimensions the problem is more complicated. The solution inside
the numerical box has not reached its far-field form and a projection
on the asymptotic state is inaccurate since the single particle and
double particle breakup wave functions live in the same regions of
space, especial near the edges of the domain. A procedure to extract
the observables with the help of surface integrals was proposed by
McCurdy, Horner and Rescigno in \cite{mccurdy2001practical}.  The
amplitude can be written as a surface integral
\begin{equation}\begin{split}
f_{l_1,l_2}(k_1,f_2) = &\frac{1}{2}\int_S \left(\tau_{k_1,l_1}(\rho_1) \tau_{k_2,l_2}(\rho_2)\bn \psi_{sc}(\rho_1,\rho_2) \right.\\
&\left. - \psi_{sc}(\rho_1,\rho_2) \bn\left(\tau_{k_1,l_1}(\rho_1) \tau_{k_2,l_2}(\rho_2)\right)\right) \cdot d\mathbf{S},
\label{eq:surfaceintegral}
\end{split}\end{equation}
where $\tau_{k_1,l_1}(\rho_1)$ is the solution of the
radial equation
\begin{equation}\label{eq:refsol}
 \left(T_1 + \frac{l_1(l_1+1)}{2 \rho_1^2}  + V_1- \frac{k_1^2}{2}\right) \tau_{k_1,l_1}(\rho) = 0.
\end{equation}
In a similar way $\tau_{k_2,l_2}(\rho)$ fits the same equation with
$T_2$, $V_2$, $l_2$ and $k_2^2$ such that $k_1^2 + k_2^2 = 2E$.  The contour of the
surface integral, \eqref{eq:surfaceintegral}, should lie in the region
where the Schr\"odinger equation is homogeneous.  In the literature
the functions $\tau_{k,l}$ are often denoted as $\phi_{k,l}$. In this
paper we have chosen a different notation to avoid confusion with
the basis functions of the oscillator representation that will be
considered in the next section.

The single differential cross section (SDCS) is then
\begin{equation}
  \frac{d\sigma}{dE_1} = \frac{8 \pi^2}{k_0^2} \frac1{k_1 k_2} \left|f(k_1,k_2)\right|^2,
\end{equation}
where $k_0$ is the momentum of the incoming wave \cite{mccurdy2001practical}.

From these amplitude formulas it is also clear why we use exterior
complex scaling rather than complex scaling. In the latter the
complete domain is rotated into the complex plane as $\rho \rightarrow
\rho e^{i\theta}$ and homogeneous Dirichlet boundary conditions are
enforced at the end of the grid.  However, after this transformation
it is very hard to extract scattering information. In contrast, the
sole purpose of ECS is to provide the correct outgoing boundary
conditions. It leaves the solution on the real part of the grid
unchanged and the scattering cross sections can be extracted.

\subsection{The $J$-Matrix Method}\label{sec:jm}

In the original $J$-Matrix method for quantum scattering
\cite{heller,heller2} the wave function is represented in terms of
some $L^2$-basis set that leads to a tri-diagonal structure of the
Hamiltonian matrix for the free-particle problem (Jacobi shape of the
matrix). This tri-diagonal structure allows the use of the complete
basis set without actually working with matrices of infinite size.  A
review of the applications of the $J$-matrix method can be found in
\cite{jmatrix}.

A popular $L^2$-basis set for calculations with short-range potentials
is the set of eigenstates of the radial harmonic oscillator. The basis states
are then generalized Laguerre polynomials multiplied by a weight
function
\begin{eqnarray}\label{eq:oscillatorfunction}
	\phi_{i,l}(\rho) &=& (-1)^i N_{i,l} b^{-3/2} \left(\frac \rho b\right)^l \nonumber \\
  &\times& \rho \,\exp\left(-\frac{\rho^2}{2 b^2}\right) L_i^{l+1/2}\left(\frac{\rho^2}{b^2}\right)	\label{osc_basis}
\end{eqnarray}
with a normalization
\begin{equation}
N_{i,l} = \sqrt{\frac{2i!}{\Gamma(i+l+3/2)}},
\end{equation}
where $i=0,1,2\ldots$ and oscillator length $b=\sqrt{\hbar/ m \omega}$
that is related to the oscillator frequency $\omega$.  Note that we
have incorporated the weight $\rho$ of integration coming from the radial coordinates in the function
\eqref{eq:oscillatorfunction} such that $\int_0^\infty
\phi_{i,l}(\rho) \phi_{j,l}(\rho) d\rho = \delta_{ij}$

Each basis function has a classical
turning point
\begin{equation}
R_{i,l} = b \sqrt{4i+2l+3}.
\end{equation}
The oscillator basis set is complete over $L^2$ and the solutions of
\eqref{eq:onedimensional}
can always be represented as a linear combination of all
oscillator states
\begin{equation}
\psi_{l}(\rho) = \sum_{i=0}^{\infty} c_{i,l} \phi_{i,l} (\rho).
\end{equation}
This representation reduces the Schr\"odinger equation to an infinite
system of linear equations for $c_{i,l}$.

As already mentioned above, the kinetic energy
operator $T$ (the free particle Hamiltonian) is a tri-diagonal matrix in the
$J$-matrix method. For the oscillator
basis the non-zero elements are
\begin{equation}
J_{i,j} = \left\{\!\!\begin{array}{c@{\quad}l}
\left(2i+l+\frac32 \right)\frac{\hbar\omega}{2} & \text{for } j=i,\\
-\sqrt{i\,\left(i+l+\frac12\right)}\frac{\hbar\omega}{2} & \text{for } j=i-1,\\
-\sqrt{(i+1)\left(i+l+\frac32 \right)}\frac{\hbar\omega}{2}& \text{for } j=i+1.
\end{array}\right.
\label{eq:kinetic_en}
\end{equation}

However, the potential energy matrix in this representation will be
dense. Thus for an accurate treatment of the problem we need to deal
with an infinitely-sized dense Hamiltonian matrix. As long as we deal
with short-range potentials only, this dense potential matrix can be
safely truncated to a finite matrix.

This relies on the fact that highly excited oscillator states (with
large index $i$) oscillate rapidly between the origin and the
corresponding classical turning point $R_{i,l}$. So, if the range of
the potential is less then $R_{i,l}$, its matrix elements will average
to negligibly small value because of annihilating oscillatory
contributions. The potential matrix can then be truncated at some
$i=N$ determined by the desired accuracy. Beyond this point, the
Hamiltonian matrix is approximated by the tri-diagonal
(asymptotic) form. We therefore refer to the dense part of the
Hamiltonian matrix corresponding to $i\leq N$ as the ``interaction
region'', and to the tridiagonal part for $i>N$ as the ``asymptotic
region''.

In the asymptotic region, the tri-diagonal structure of the matrix leads
to a simple three-term recurrence relation for the oscillator expansion
coefficients of the solution:
\begin{equation}
J_{i,i-1}c_{i-1} + (J_{i,i} - E)c_{i} + J_{i,i+1} c_{i+1} = 0  \quad \forall  i \ge N \label{eq:recrel}
\end{equation}
Since this is a second order recurrence relation the $\{c_i\}$ can be
obtained as a linear combination of two linearly independent
fundamental solutions. In the $J$-matrix method the regular $b_{i,l}$ and
the irregular $n_{i,l}$ are chosen as fundamental solutions that can
be easily obtained from the explicit form of the matrix $J$, eq.
(\ref{eq:kinetic_en}),
\begin{equation}
c_{i,l} = b_{i,l} + t\, n_{i,l}  \quad \forall i \ge N.
\end{equation}
These fundamental solutions of the recurrence relation have a corresponding coordinate-space solution
\begin{equation}
\psi_{l}(\rho) = \mathcal{B}_{l}(\rho) + t\, \mathcal{N}_{l}(\rho),
\end{equation}
where $\mathcal{B}_{l}(\rho) = \sum_{i=0}^{\infty} b_{i,l}
\phi_{i,l}(\rho)$, $\mathcal{N}_{l}(\rho) = \sum_{i=0}^{\infty}
n_{i,l} \phi_{i,l}(\rho)$ are usually referred to as Bessel-like and
Neumann-like respectively, due to their asymptotic behavior. In this
context the coefficient $t$ in this linear combination corresponds to the scattering $t$-matrix.

The full solution is then
\begin{equation}
\psi_{l}(\rho) = \chi_{l}(\rho) + \mathcal{B}_{l}(\rho) + t\, \mathcal{N}_{l}(\rho),
\end{equation}
where the function $\chi_{l}(\rho) = \sum_{i=0}^{\infty} c^{0}_{i,l}
\phi_{i,l}(\rho)$ is nonzero only in the interaction
region. This leads to
\begin{equation}
c_{i,l} = \left\{ \begin{array}{c@{\quad\text{when }}l} c^0_{i,l} + b_{i,l} + t\, n_{i,l} & i<N \\ b_{i,l} + t\, n_{i,l} & i \geq N \end{array}. \right.
\end{equation}

With this form of the solution we reduce the set of unknowns to $\{c^0_{i,l}, t\}$ and the linear system to be solved has $N+1$ dimensions. Solving the linear system we obtain simultaneously the wave function of the system, $\{ c^0_{i,l}\}$, and the scattering information, $t$.

\subsection{Introduction of the JM-ECS method}
\label{sec:jmecs}
For multi-particle scattering and reaction (breakup) problems it is
very hard to obtain an explicit form of the asymptotic solutions, also
in the oscillator representation. It is then a natural approach to
avoid such explicit forms by introducing the ECS transformation in the
asymptotic JM region by enforcing outgoing wave boundary conditions.

A formal introduction of the ECS within the oscillator basis
representation is hard for several reasons. First, applying the
ECS coordinate transformation \eqref{eq:ecstrafo} to oscillator
functions destroys the orthogonality as well as the tri-diagonal
structure of the kinetic energy matrix. Second, the convergence of the
oscillator basis will be strongly affected. Most probably, the wave
function within the absorbing layer will be poorly reproduced even by
a huge truncated basis set. Finally, the coordinate transformation
makes an analytical calculation of matrix elements difficult.

An alternative approach is to combine the grid and the oscillator
representation and represent the wave function in the asymptotic
region with finite differences. This becomes possible due to a
specific property of the oscillator basis. For highly excited
oscillator states the oscillator expansion coefficients are related to
the values of the wave function on the grid of classical turning
points \cite{Vanroose2001}:
\begin{equation}
\begin{split}
c_{n,l} &= \int_0^\infty \phi_{i,l}(\rho) \psi_{l}(\rho) d\rho \\
&\approx b \sqrt{2 /R_{n,l}} \left[ \Psi_l(R_{n,l})+\mathcal{O}\left(\frac{1}{R_{n,l}}\right) \right].
	\label{eq:as_relation}
\end{split}
\end{equation}
This allows to couple the oscillator representation of the wave
function with the coordinate-space representation in the high-$n$
region.  A similar argument has been used for
building the so-called modified $J$-matrix method (MJM) \cite{Vanroose2001}.

\subsubsection{A hybrid  representation of the wave function}
In the hybrid JM-ECS method we represent our one-dimensional 
wave function as a vector $\mathsf{\Psi}$ in $\mathbb{C}^{n+m}$,  where
\begin{equation} \label{eq:hybridvector}
\begin{split}
     \mathsf{\Psi} = (&c_0,c_1, \ldots, c_{n-1},\\
     &\psi(R_{n}),\psi(R_{n}\! +\! h),\ldots,\psi(R_{n}\! +\! (m\!-\!1)h) ).
\end{split}
\end{equation}
The first $n$ elements represent the wave function in the
\textit{interaction region} in the oscillator representation, while
the remaining $m$ elements represent the wave function in the
\textit{asymptotic region} on an equidistant grid that starts at
$R_n$, the $n$-th classical turning point, and runs up to $R_n +
(m-1)h$ with a grid distance
\begin{equation}
  h = R_{n}-R_{n-1}.
  \label{eq:gridsize}
\end{equation}
We assume that the matching point, which connects the oscillator to
finite difference representation, corresponds to an large index $n$
such that the asymptotic formula for the
expansion coefficient, \eqref{eq:as_relation}, can be applied.

Again, the kinetic energy operator in this hybrid representation is
tridiagonal since it is tridiagonal both in finite-differences and
oscillator representations. One should only be careful near the
matching point between both representations so that the asymptotic
formula \eqref{eq:as_relation} is a sufficiently good approximation for
representing the solution.

To obtain the kinetic energy in the final point of the oscillator
representation, we use the tridiagonal kinetic energy formula
\eqref{eq:kinetic_en}. It involves a recurrence relation connecting
the three terms $c_{n-2}$, $c_{n-1}$ and $c_{n}$. The latter,
the coefficient $c_n$, is unknown. Only $\psi(R_n)$ is available.  Using
the asymptotic relation \eqref{eq:as_relation}, however, we can
calculate the required matrix element as follows:
\[\begin{split}
 (Jc)_{n-1} = &J_{n-1,n-2} c_{n-2} + J_{n-1,n-1} c_{n-1}\\& + J_{n-1,n} b \sqrt{2/ R_{n}} \psi_l(R_{n}).
\end{split}\]

To calculate the kinetic energy in the first point of the finite
difference grid, the second derivative of the wave function has to be
known.  To approximate the latter with a finite difference formula,
one needs the wave function in the grid points $R_{n-1}$, $R_{n}$ and $R_n + h$. We again
apply \eqref{eq:as_relation} to obtain $\psi(R_{n-1})$ in terms of $c_{n-1}$:
\[
\psi''(R_n) = \frac{c_{n-1}/( b \sqrt{2/ R_{n-1}}) - 2 \psi(R_{n}) + \psi(R_{n}+h)}{h^2}.
\]

\begin{figure}[ht]
\hspace{10pt} Oscillator \hspace{30pt} Finite Differences
$$
\overbrace{\vphantom{\rule[5pt]{5pt}{5pt}}c_{n-2} \;\; \quad c_{n-1} \qquad \Psi(R_n)}^{T c_{n-1}} \quad \Psi(R_{n}+h) \makebox[0pt][r]{$\underbrace{ \phantom{c_{n-1} \qquad \Psi(R_n) \quad \Psi(R_{n+1}) \vphantom{\rule[-20pt]{20pt}{20pt}} }}_{\Psi''(R_n)}$}
$$
\ifx\JPicScale\undefined\def\JPicScale{0.6}\fi
\unitlength \JPicScale mm
\begin{picture}(120,30)(-1,-32)
\linethickness{0.3mm}
\put(0,20){\line(1,0){120}}
\linethickness{0.3mm}
\put(90,20){\circle*{3}}

\linethickness{0.3mm}
\put(20,20){\circle{3}}

\linethickness{0.3mm}
\put(40,20){\circle{3}}

\linethickness{0.3mm}
\put(67,20){\circle*{3}}

\linethickness{0.1mm}
\put(53,-6){\line(0,1){50}}
\end{picture}
\vspace{-30pt}
 \caption{This figure illustrates how the kinetic energy matrix
   elements are calculated in the last point $R_{n-1}$ of the
   oscillator representation and in the first point $R_n$ of the
   finite difference representation.  To calculate $T$ applied on a
   solution vector we need to translate the oscillator representation
   to the grid and vice versa. \label{fig:coupling}}
\end{figure}
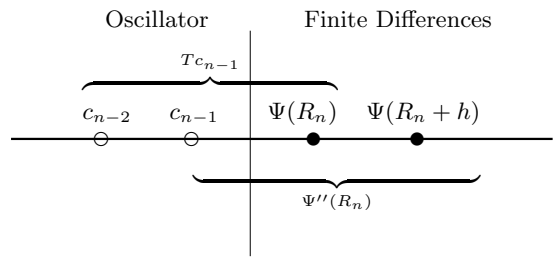

The coupling between both representations around the matching point is
sketched in Fig. \ref{fig:coupling}, together with the terms involved
to determine the correct matching.

The structure of the matrix representation of the kinetic operator
around the matching point is then
\begin{widetext}
\begin{equation}
\left(\begin{array}{cccccc}
 \ddots  & \ddots &\\
   J_{n-1,n-2} & J_{n-1,n-1}    &  J_{n-1,n} b \sqrt{2/R_n}\\
 &   1/(h^2 b \sqrt{2 /R_{n-1}})   & -2/h^2   & 1/h^2\\
 &&\ddots&\ddots\\
\end{array}\right)
\left(\begin{array}{c}
\vdots\\
c_{n-1}\\
\psi(R_n)\\
\vdots\\
\end{array}\right).
\end{equation}
\end{widetext}
	
The representation of the potential operator in the hybrid JM-ECS
method is more complex. Similar to the JM method the potential matrix
in the interaction region, covered by the oscillator representation,
is dense. Therefore, the full hybrid potential matrix will be dense in
the interaction region and diagonal in the asymptotic finite-differences
region. It is clear that the computational complexity of the problem
is determined by the size of the interaction region because of
the large number of potential matrix elements that needs to be
calculated. Shifting the matching point further outward will therefore
strongly affect the computation time in a negative way.  Increasing the
finite-differences asymptotic part will have almost no effect on the
latter because of the diagonal potential.

The introduction of the outgoing wave boundary condition is now
straightforward by extending the finite difference grid with an ECS
contour as explained in section \ref{sec:ecs}. We use $m$ grid
points to cover both the real part and the ECS part of the finite
difference grid.

Similar to the finite difference representation \eqref{eq:twoDECS} one
can easily construct a representation for a two-dimensional wave
function.
Instead of the
vector form \eqref{eq:hybridvector}, the solution can be represented
by a matrix $\mathsf{\Psi} \in \mathbb{C}^{(n+m)\times(n+m)}$ that has
the following structure
\begin{widetext}
\begin{equation}
\mathsf{\Psi} = 
\left(\begin{array}{lll|lll}
c_{0\,0}      & \ldots &c_{0n-1}    &  d_{0\,n}            & \ldots &d_{0{n+m}}\\
c_{1\,0}      & \ldots &c_{1n-1}    &  d_{1\,n}            & \ldots &d_{1{n+m}}\\
c_{n-1\,0}    & \ldots &c_{n-1n-1}  &  d_{n-1\,n}          & \ldots   \\
\hline
d_{n\,0}      & \ldots &d_{n\,n-1}    &  \psi(R_n,R_n)     & \ldots &\psi(R_n,R_{n+m})\\
d_{n+1\,0}    & \ldots &d_{n+1\,n-1}  &  \psi(R_{n+1},R_n) & \ldots &\psi(R_{n+1},R_{n+m})\\
\hdots      &        & &  \hdots  &                                                   \\
d_{n+m\,0}    & \ldots &d_{n+m\,n-1}  &  \psi(R_{n+m},R_n) & \ldots &\psi(R_{n+m},R_{n+m})\\
\end{array}\right).
\end{equation}
\end{widetext}

\begin{figure}[ht]
\includegraphics[width=0.9\linewidth]{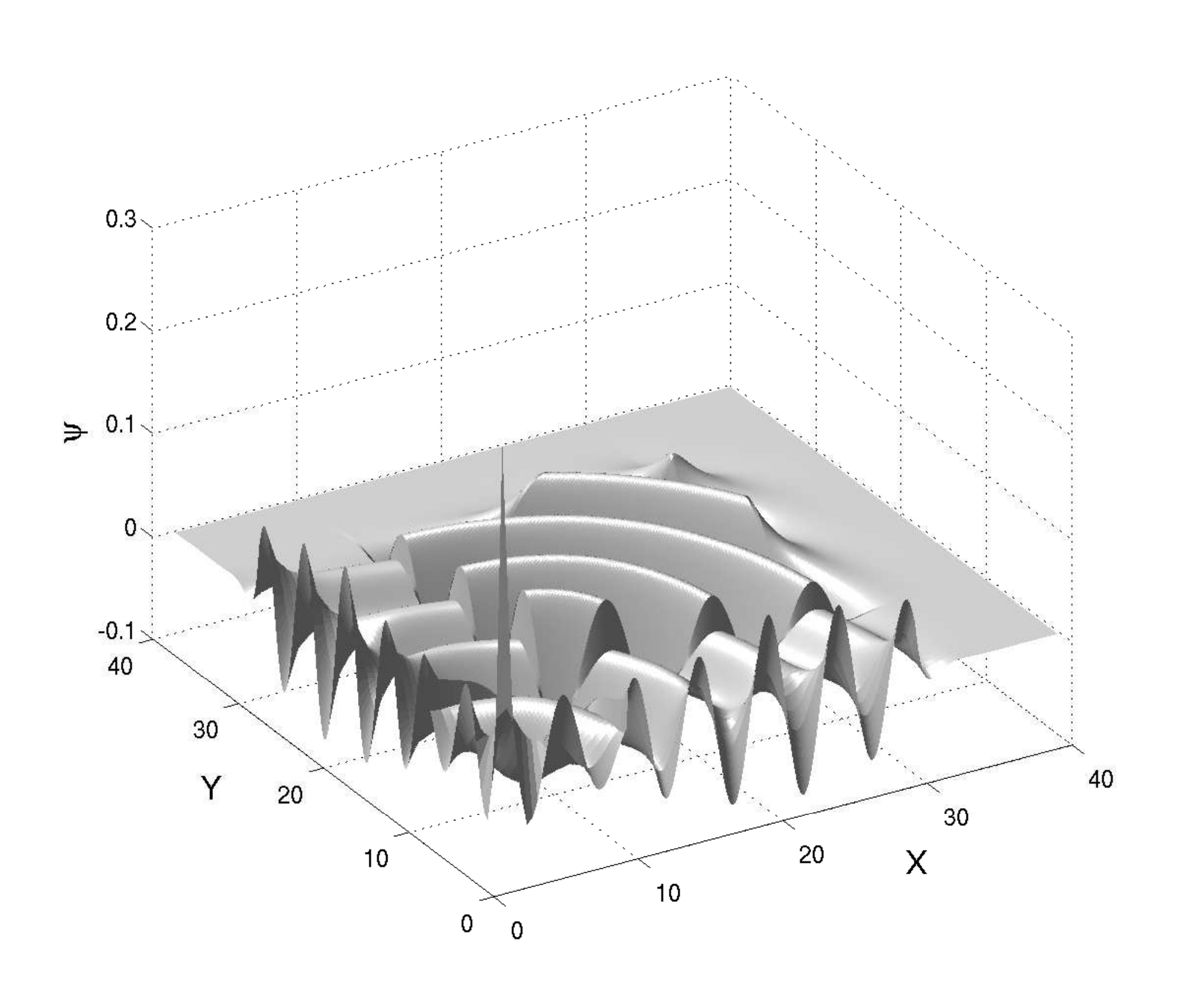}
\caption{The real part of the 2D wave function of the Three-particle
  s-wave scattering problem of \eqref{eq:2d_1} in the hybrid approach
  with 40 oscillator states and the oscillator parameter $b=0.8$. The
  values of $X$ and $Y$ in the oscillator region are chosen as the
  corresponding classical turning points to show the similar behavior
  of the wave function in different representations.  Compare with
  Fig.\ \eqref{fig:sample_2d}. Note the jump in the wave function near
  the J-matrix and finite difference boundary. This jump clearly
  indicates the different regions in the
  representation.}\label{fig:hyb_surf}
\end{figure}

An example of this wave function is presented in Fig.~\ref{fig:hyb_surf}. The spatial wave function can then, depending on the region in the two-dimensional domain, be written as follows. For $\rho_1,\rho_2 < R_n$ one
has
\begin{equation}
  \psi(\rho_1,\rho_2) = \sum_{i,j=0}^{n-1} \mathsf{\Psi}_{ij} \phi_i(\rho_1)\phi_j(\rho_2), 
\end{equation}
for $\rho_1 < R_n$ and $k \ge n$ one writes
\begin{equation}
  \psi(\rho_1,R_l) =  \sum_{i}^{n-1} \mathsf{\Psi}_{il} \phi_i(\rho_1)
\end{equation}
and for $\rho_2 < R_n$ and $l \ge n$,
\begin{equation}
  \psi(R_k,\rho_2) =  \sum_{j}^{n-1} \mathsf{\Psi}_{kj} \phi_j(\rho_2)
\end{equation}
and, finally, for $k,l \ge n$
\begin{equation}
  \psi(R_k,R_k) =  \mathsf{\Psi}_{kl}.
\end{equation}

The Hamiltonian of a 2D problem is again constructed as a Kronecker
product of the 1D Hamiltonians and a two-body potential.
The locations of non-zero matrix elements are the
same as in the Kronecker product of the two two-particle potential
matrices. This leads to a number of dense blocks distributed over a
large sparse matrix. In this case the computational complexity is also determined by the
size of the finite-differences part, as it not only extends the
diagonal, but also increases the number of dense blocks.

\subsubsection{Extracting Scattering information from a wave function in a hybrid representation}\label{sec:extracthybrid}
As already mentioned at the beginning of this section, the extraction
of observables is not always straightforward in the JM method. This
mainly comes from the fact that in the original formulations one has to solve the scattering
problem and define the scattering parameters simultaneously.

In the hybrid JM-ECS approach in one dimension, once a solution is
obtained, we still need to extract the observables. Since the wave
function is represented in the asymptotic region with a
finite-difference representation, we can use a standard Wronskian
technique to extract the scattering amplitude (see section
\ref{sec:extract}).

\begin{figure}
\begin{center}
\psfrag{L}{$x^2$}
\includegraphics[width=0.9\linewidth]{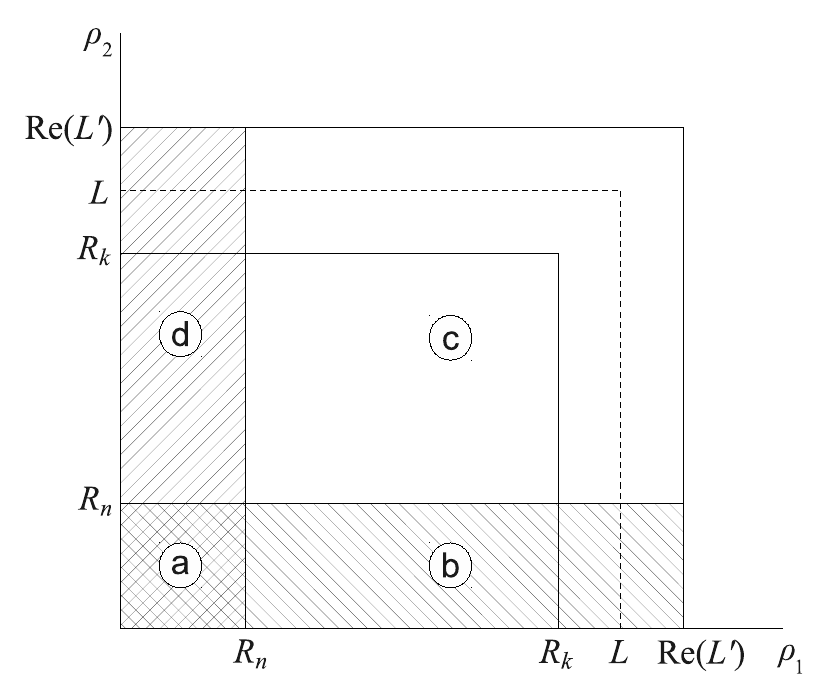}
\end{center}
\caption{There are four region in the hybrid representation. In region
  a) $\rho_1, \rho_2 < R_n$ and the wave function is a sum over
  product of two oscillator functions. In b) and d) one one
  coordinate, respectively $\rho_1 < R_n$ and $\rho_2 < R_n$, is
  represented in the oscillator state. In region c) we use finite
  difference for both coordinates.  The amplitude is calculated with
  the help of a surface integral along the path that is indicated by
  the solid line a distance $R_k$ away from the axes.  Note that
  between $L$ and Re$(L^\prime)$ the grid is complex valued because of
  ECS. \label{fig:2ddomain}}
\end{figure}

Obtaining the scattering information from the solution in two
dimensions requires the surface integral of section
\ref{sec:extract}. This approach still needs to be elaborated in the
hybrid representation.
We consider for this discussion a contour $S$ of the surface integral
\eqref{eq:surfaceintegral} that is piecewise parallel to one of the axes.
This is illustrated in Fig.~\ref{fig:2ddomain}. We first integrate
$\rho_2$ from 0 to a value $R_k \in[a,L]$ while fixing
$\rho_1=R_k$. The point $R_k$ is located on the finite difference grid
in the region where the equation is homogeneous. Typically $R_k$ is
chosen a few grid points before the end of the real part of the
grid. This procedure leads to a line integral parallel to the $\rho_2$
axis. We denote it as $I_2$.

The second part of the surface integral integrates $\rho_1$ from $R_k$
to 0 while keeping $\rho_2$ equal to $R_k$. This is a line integral
parallel to the $\rho_1$ axis and is denoted as $I_1$.

%
%
The amplitude, \eqref{eq:surfaceintegral}, now consists of two parts
\begin{equation}
f_{l_1,l_2}(k_1,k_2) = I_1 + I_2,
\end{equation}
where 
\begin{equation}\label{eq:int1}
\begin{split}
I_1 = \frac{1}{2} 
\int_0^{R_k} &\left( \tau_{k_1,l_1}(\rho_1) \tau_{k_2,l_2}(R_k)\frac{\partial \psi_{sc}(\rho_1,R_k)}{\partial \rho_2} \right.\\
&\left. - \psi_{sc}(\rho_1,R_k) \tau_{k_1,l_1}(\rho_1) \frac{\partial\tau_{k_2,l_2}(R_k)}{\partial \rho_2}\right)  d\rho_1
\end{split}
\end{equation}
 and 
\begin{equation}\label{eq:int2}
\begin{split}
I_2 = \frac{1}{2} 
\int_0^{R_k} &\left(\tau_{k_1,l_1}(R_k) \tau_{k_2,l_2}(\rho_2)\frac{\partial \psi_{sc}(R_k,\rho_2)}{\partial \rho_1} \right.\\
&\left. - \psi_{sc}(R_k,\rho_2) \tau_{k_2,l_2}(\rho_2) \frac{\partial\tau_{k_1,l_1}(R_k)}{\partial \rho_1} \right) d\rho_2.
\end{split}
\end{equation}
This is in fact a surface integral of an in-product of two
vector-valued functions. We can reverse the bounds on $I_1$ without
switching the sign of the argument, since the surface vector also
changes direction.

It is important to note that $I_1$ probes the 2D wave function
$\psi_{sc}(\rho_1,\rho_2)$ where it is either represented as a sum
over oscillator states in the first coordinate and a finite difference
in the second or as a finite difference in both coordinates.  The
integral never probes the region where both coordinates are
represented in the oscillator representation (see region (a) in Fig.
\ref{fig:2ddomain}).

To arrive at a practical expression for the amplitude, we split $I_1$ into $\int_0^{R_n}(\ldots) + \int_{R_n}^{R_k} (\ldots)$ into an integration
from 0 to $R_n$, the part covered by the oscillator representation,
and an integration from $R_n$ to $R_k$, covering the grid.

In order to calculate the integrals we require the solutions
$\tau_{k_1,l_1}(\rho)$ and $\tau_{k_2,l_2}(\rho)$ of equation
\eqref{eq:refsol}. We solve these equations also in the hybrid
representation. The solutions are vectors $t(k_1)$ and
$t(k_2) \in \mathbb{C}^{(n+m)}$ representing the numerical
solution of $\tau_{k_1,l_1}(\rho)$ and $\tau_{k_2,l_2}(\rho)$,
respectively, with $k_1^2 + k_2^2 = E$. For $\rho_1 \in [0,R_n]$ the solution is a finite sum
\[
\tau_{k_1,l_1}(\rho_1) = \sum_{i}^{n-1} t(k_1)_i
\phi_i(\rho_1)
\]
and, similarly, the 2D wave function in region (b) of Fig. \eqref{fig:2ddomain} equals
$\psi_{sc}(\rho_1,R_k) =
\sum_{i=0}^{n-1}\mathsf{\Psi}_{ik}\phi_i(\rho_1)$.

The calculation of the amplitude requires the first order
derivatives $\partial \tau_{k_1,l_1}(R_k)/\partial \rho_1$ and $\partial
\psi_{sc}(\rho_1,R_k)/\partial \rho_2$. These will be approximated by
central differences from $t(k_1)$ and $\mathsf{\Psi}$. We define
\begin{equation}
  t_k^\prime(k_1) = \frac{t(k_1)_{k+1}-t(k_1)_{k-1}}{2h},
\end{equation}
and similarly for $t^\prime(k_2)_k$. We also define
\begin{equation}
  \mathsf{\Psi}^{\prime,1}_{ik} = \frac{\mathsf{\Psi}_{i+1\,k}-\mathsf{\Psi}_{i-1\,k}}{2h}
\end{equation}
and \begin{equation}
  \mathsf{\Psi}^{\prime,2}_{ik} = \frac{\mathsf{\Psi}_{i\,k+1}-\mathsf{\Psi}_{i\,k-1}}{2h}.
\end{equation}

So the first term in the integral of $I_1$ becomes
\begin{equation}
\begin{split}
\int_0^{R_n}& \left(-\tau_{k_1,l_1}(\rho_1) \tau_{k_2,l_2}(R_k)\frac{\partial \psi_{sc}(\rho_1,R_k)}{\partial \rho_2} \right.\\
&\left. + \psi_{sc}(\rho_1,R_k) \tau_{k_1,l_1}(\rho_1) \frac{\partial \tau_{k_2,l_2}(R_k)}{\partial \rho_2}\right)  d\rho_1\\
 = \int_0^{R_n} &\left[ -\left(\sum_{i=0}^{n-1} t_i(k_1) \phi_i(\rho_1)\right)t_k(k_2) \left(\sum_{j=0}^{n-1} \mathsf{\Psi}^{\prime,2}_{jk} \phi_j(\rho_1)\right) \right.\\
&\!\!\!\!\!\!\! \left. +\left( \sum_{j=0}^{n-1} \mathsf{\Psi}_{jk} \phi_j(\rho_1)\right) \left(\sum_{i=0}^{n-1}t_i(k_1)\phi_i(\rho_1)\right) t_k^\prime(k_2)\right]  d\rho_1,
\end{split}
\end{equation}
where we have used that $\tau_{k_2,l_2}(R_k) = t_k(k_2)$ since the
vector $t(k)$ represents the solution in the hybrid representation. In
the next step we reorder the sum and the integral and find
\begin{equation}
\begin{split}
 = - t_k(k_2) &\sum_{i=0}^{n-1}\sum_{j=0}^{n-1}  t_i(k_1) \mathsf{\Psi}^{\prime,2}_{jk} \int_0^{R_n} \phi_i(\rho_1) \phi_j(\rho_1) d\rho_1\\
&  +  t_k^\prime(k_2) \sum_{j=0}^{n-1}\sum_{i=0}^{n-1} t_i(k_1) \mathsf{\Psi}_{jk} \int_0^{R_n} \phi_j(\rho_1)\phi_i(\rho_1) d\rho_1\\
  \approx -t_k(k_2) &\sum_{i=0}^{n-1} t_i(k_1) \mathsf{\Psi}^{\prime,2}_{ik}  + t^\prime_k(k_2) \sum_{i=0}^{n-1} t_i(k_1)\mathsf{\Psi}_{ik} \\
& + \mathcal{O}(\max_{i,j<n}|\int_{R_{n}}^{\infty} \phi_i \phi_j|).
\end{split}
\end{equation}
We have used that $\int_0^{R_n} \phi_i \phi_j = \delta_{ij} -
\int_{R_n}^{\infty}\phi_i \phi_j$.

The second term in the integral $I_1$ covers the grid and it is
approximated using Simpson's rule
\begin{equation}
\begin{split}
&\int_{R_n}^{R_k}\left( -\tau_{k_1,l_1}(\rho_1)
\tau_{k_2,l_2}(R_k)\frac{\partial}{\partial \rho_2}
\psi_{sc}(\rho_1,R_k) \right.\\ &\left. + \psi_{sc}(\rho_1,R_k)
\tau_{k_1,l_1}(\rho_1) \frac{\partial}{\partial
  \rho_2}\tau_{k_2,l_2}(R_k)\right) d\rho_1 \\
 &= - t_k(k_2)
\sum_{i=n}^{k} t_i(k_1) \mathsf{\Psi}^{\prime,2}_{ik} w_i +
t^\prime_k(k_2)\sum_{i=n}^{k} \mathsf{\Psi}_{ik} t_i(k_1) w_i,
\end{split}
\end{equation}
where $w_i$ is the weight of integration. The weight has a value $w_i=h$, for all $i$ except at
$i=n$ and $i=k$, the end points of integration, where it equals $w_i=h/2$.
Combining both parts $I_1$ and $I_2$ one obtains
\begin{equation}
\begin{split}\label{eq:finalamp}
 &2  f_{l_1,l_2}(k_1,k_2) =\\
&  t_k(k_1) \sum_{i=0}^{k} t_i(k_2) \mathsf{\Psi}^{\prime,1}_{ki} w_i  -  t^\prime_k(k_1)\sum_{i=0}^{k} \mathsf{\Psi}_{ki} t_i(k_2) w_i\\
& - t_k(k_2) \sum_{i=0}^{k} t_i(k_1) \mathsf{\Psi}^{\prime,2}_{ik} w_i  +  t^\prime_k(k_2)\sum_{i=0}^{k} \mathsf{\Psi}_{ik} t_i(k_1) w_i,
\end{split}
\end{equation}
where 
\begin{equation}
  w_i = \begin{cases}
    1   \quad  &\text{if}  \quad i < n,\\
    h/2  \quad  &\text{if} \quad i = n,\\
    h  \quad  &\text{if} \quad n < i < k,\\
    h/2  \quad  &\text{if} \quad i = k.
  \end{cases}
\end{equation}
The factor 2 in \eqref{eq:finalamp} comes from the surface integral \eqref{eq:surfaceintegral}. 
\section{Numerical results} \label{results}
To illustrate and benchmark the proposed hybrid JM-ECS method, we
consider two model problems that are derived from a two-body and a
three-body problem.  When these systems are expressed in spherical
coordinates around the center of mass and expanded in spherical
harmonics one typically ends up with a large system of coupled radial
equations, known as the partial wave equations.  These partial waves
can either be coupled 1D or coupled 2D problems.  We choose to benchmark our methods
to model problems formulated as an uncoupled partial wave problem.

The two-body problem reduces to a one-dimensional radial problem
with the form
given by \eqref{eq:onedimensional}.

Similarly a two-dimensional radial problem is derived from a
three-body problem and is, expressed as:
\begin{equation}\label{eq:twodimensional}
\left\{\begin{split}
&\left(T_1 +V_1(\rho_1) + \frac{l_1(l_1+1)}{2\rho_1^2}  + T_2  + V_2(\rho_2) + \frac{l_2(l_2+1)}{2\rho^2_2} \right.  \\
&\hspace{0.7cm}\left. + V_{12}(\rho_1,\rho_2) - \frac{k^2}{2}\right) \psi_{sc}(\rho_1,\rho_2) \\
&\hspace{0.2cm}= - \frac1{\sqrt{k_0}}\left(\left[V_2(\rho_2) + V_{12}(\rho_1,\rho_2)\right] \varphi_{l_1,0}(\rho_1) \hat{j}_{l_2}(k_0\rho_2) \right. \\
&\hspace{1.5cm}+\left.\left[V_1(\rho_1) + V_{12}(\rho_1,\rho_2)\right] \varphi_{l_2,0}(\rho_2) \hat{j}_{l_1}(k_0\rho_1) \right),\\
&\psi_{sc}(0,\rho_2)=0 \,\,\forall \rho_1 > 0 \quad \text{and} \quad \psi_{sc}(\rho_1,0)=0 \,\,\forall \rho_2\\
&\text{with outgoing wave boundary conditions for}\\
&\rho_1\rightarrow  \infty \quad \text{or} \quad \rho_2\rightarrow \infty,
\end{split}
\right.
\end{equation}
where $\rho_1$ and $\rho_2$ are two radial coordinates representing the
distances of the first and second particle to the center of
the coordinate system. The two body potential is
$V_{12}(\rho_1,\rho_2)$. The angular momenta $l_1$ and $l_2$ are
non-negative integers.

The total wave function is the sum of an incoming wave and a scattered
wave. If we model an impact ionization problem, for example, the
incoming wave will be a product of the target in the ground state, $\varphi_{l_1,0}$ and
the incoming wave, $j_{l_1}(k_0 \rho_2)$, of the second particle. Taking into account
symmetrization this is
$$
\frac1{\sqrt{k_0}}\left(\varphi_{l_1,0}(\rho_1)\hat{j}_{l_2}(k_0\rho_2) + \varphi_{l_2,0}(\rho_2)\hat{j}_{l_1}(k_0\rho_1) \right),
$$ 
where
 $\varphi_{l_1,0}(\rho_1)$ is an eigenstate of the operator
 $(T_1+V_1+l_1(l_1+1)/2 \rho_1^2)$ with energy $E_0$. The incoming
 wave $\hat{j}_{l_1}(k_0\rho_1)$ has momentum $k_0$ such that
 $k_0^2/2 + E_0 = k^2/2$.

Note that the method we propose is not only applicable to impact
problems but also to other breakup situations. Then the right hand side in
\eqref{eq:twodimensional} can be replaced by any other driving
term. In photo-ionization, for example, the right hand side is the
dipole operator $\vec{\mu}$ applied to the ground state of the two
particle system $\varphi(\rho_1,\rho_2)$.

\subsection{One dimensional phase shift}

\begin{figure}[ht]
\includegraphics[width=0.9\linewidth]{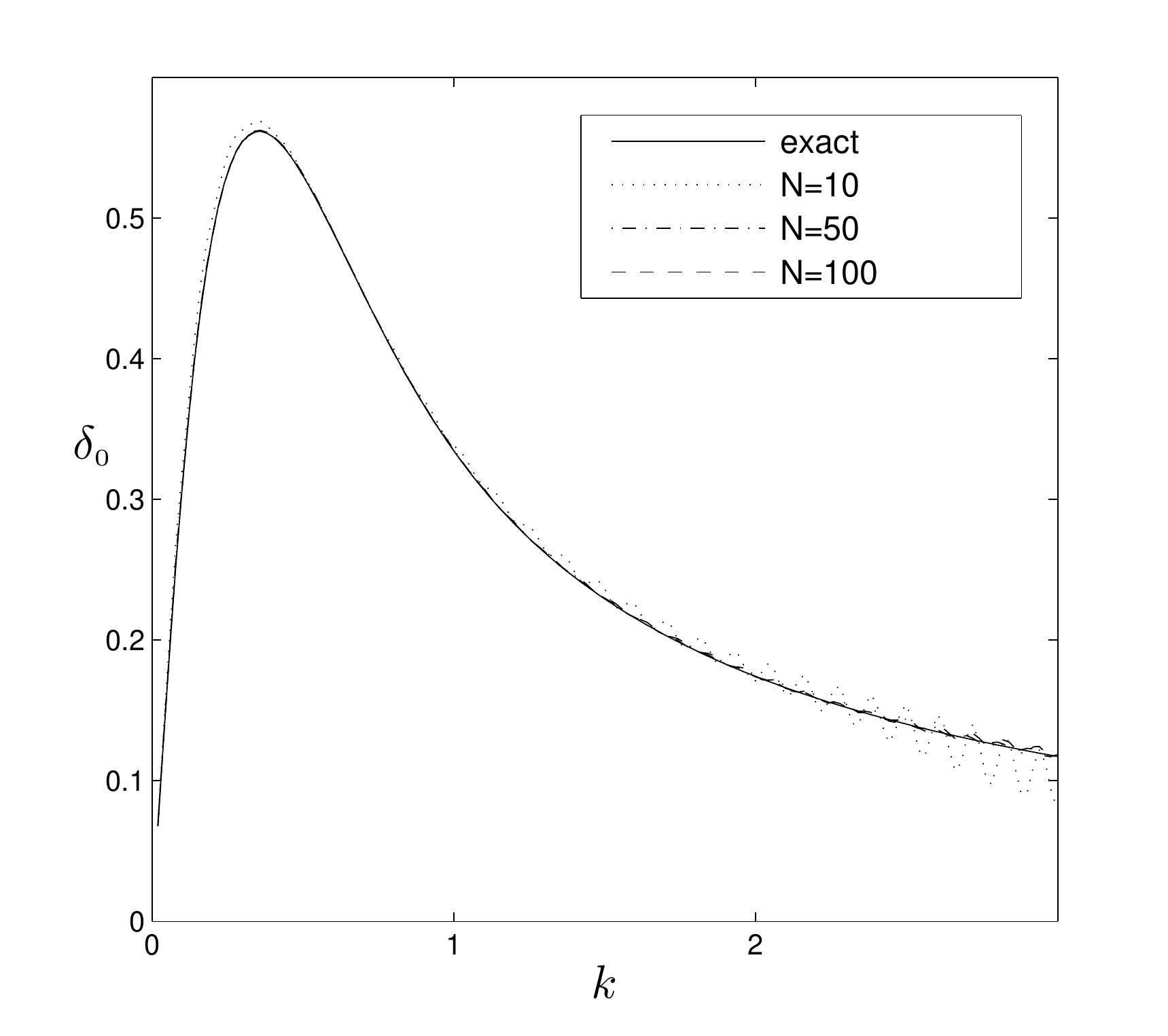}
 \caption{ $s$-wave phase shift for a Gaussian potential ($V_0 = 0.2$,
   $r_0 = 2$) as a function of wave number $k$ for the problem of
   \eqref{example_1d}. Shown are the hybrid JM-ECS results for different
   matching points $N$, and the VPA (exact) result. The oscillator length is $b=0.3$.}
 \label{fig:ph0}
\end{figure}

\begin{figure}[ht]
\includegraphics[width=0.9\linewidth]{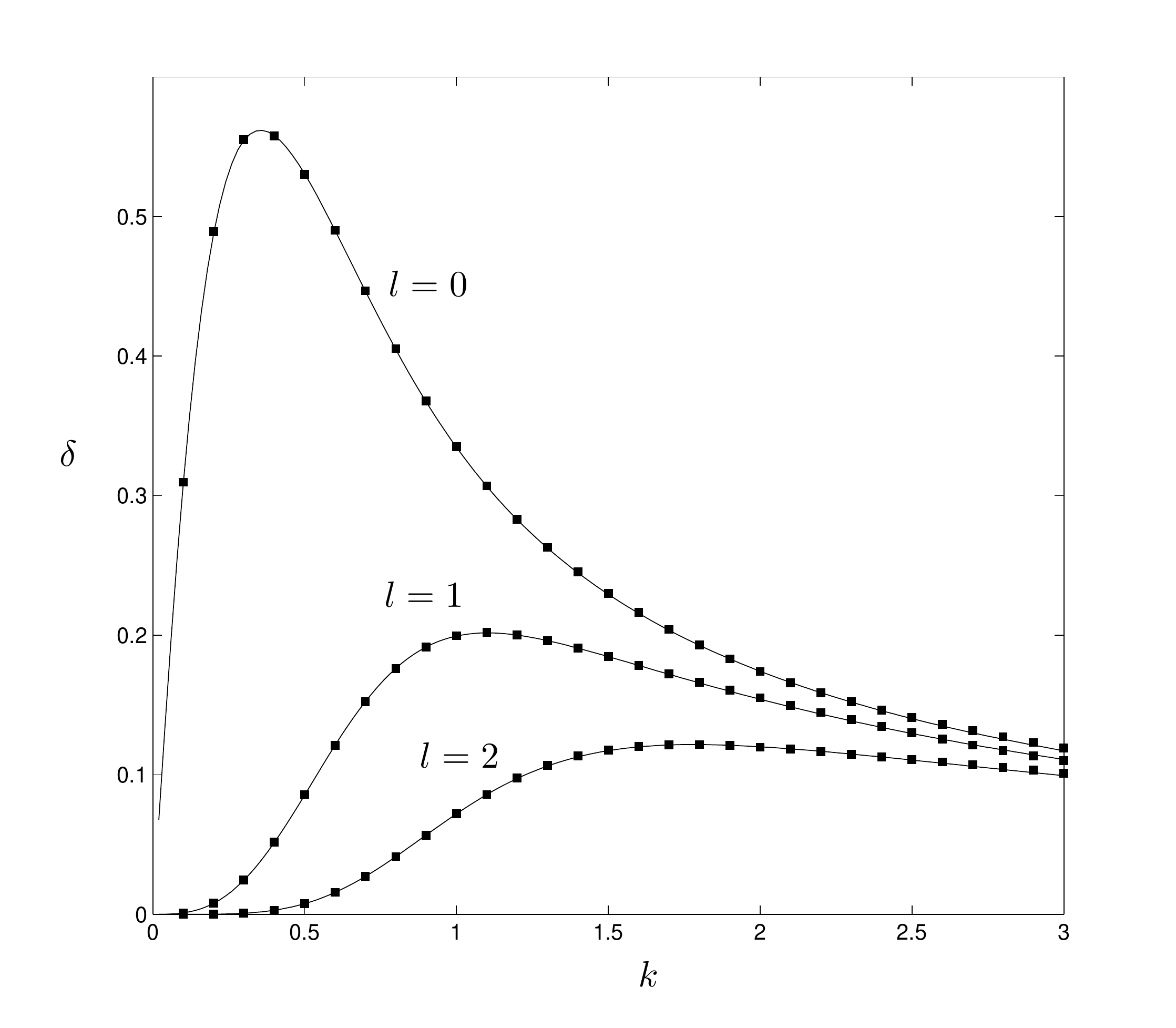}
 \caption{Phase shifts for a Gaussian potential ($V_0 = 0.2$, $r_0 =
   2$) of equation \eqref{example_1d} and different angular momenta obtained with JM-ECS (squares), compared to VPA
   results (solid lines). For all results $b=0.3$ and $N=50$.}
  \label{fig:ph012}
\end{figure}

To validate the proposed hybrid JM-ECS technique we 
first solve the one-dimensional scattering
problem with an attractive Gaussian potential:
\begin{equation}
  \left(-\frac12 \frac{d^2}{dx^2} + \frac{l(l+1)}{2 x^2} - V_0 \mathrm{e}^{x^2/r_0^2} - \frac{k^2}{2}\right) \psi(x) = 0,   \label{example_1d} 
\end{equation}
where the parameters of the potential were chosen as $V_0=0.2$ and
$r_0=2$.  We determine the elastic scattering phase shift as a
function of the momentum of the incoming particle.  To estimate the
accuracy of the hybrid results we compare the phase shift to the one
obtained with the Variable Phase Approach (VPA) \cite{calogeroVPA}, which yields the exact result with very high
accuracy for this simple problem. In Fig.~\ref{fig:ph0} we display the results for oscillator
length $b=0.3$ and increasing
matching point $N$; this corresponds to an increasing size of the
interaction region, and thus an increasing size of the truncated
oscillator basis. It is clear from this figure that the results
strongly depend on the choice of the matching point for a required
accuracy. In Fig.~\ref{fig:ph012} we display the results for different
values of the angular momentum $l$, all seen to be of qualitatively
comparable accuracy.

\begin{figure}[ht]
\includegraphics[width=0.9\linewidth]{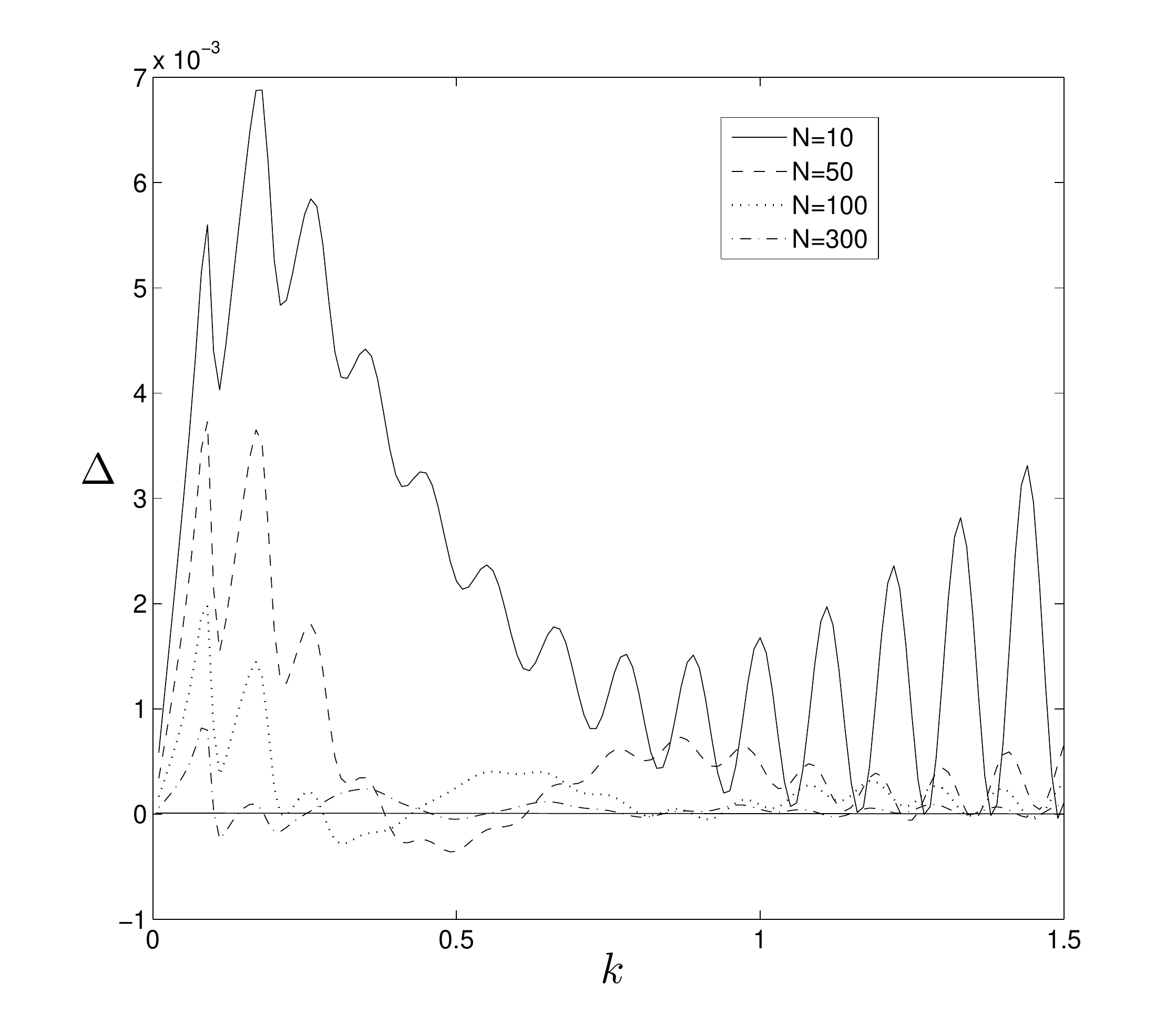}
 \caption{The absolute error in phase shift for different sizes of the
   oscillator basis (b=0.3, l=0) for model problem of fig. \ref{fig:ph0} and \ref{fig:ph0}}
 \label{fig:error1d}
\end{figure}

To quantify the accuracy we display an error plot in
Fig.~\ref{fig:error1d} where one notices an increasing accuracy with
increasing size of the oscillator basis in the interaction region.
The error is seen to decrease in terms of $N$ in both the small- and
large-energy region. For large energies the error is mainly caused by
the finite-differences approximation to the second derivative. At small energies the
error mainly comes from an inaccurate approximate relation
(\ref{eq:as_relation}) which improves as the size of the oscillator
basis increases.

In table~\ref{tab:error1d} we display the error value for different
sizes $N$ of the oscillator basis and for different partial waves. For
each partial wave we consider the $k$-value where the error is maximal

\begin{table}[ht]
\begin{center}
\renewcommand{\arraystretch}{1.1}
\renewcommand{\tabcolsep}{3mm}
\begin{tabular}{c|c|c|c}
\hline
$N$ & \multicolumn{3}{c}{$|\Delta|$ ($\times 10^{-3}$)}\\
\cline{2-4}
 & $l=0$  & $l=1$  & $l=2$  \\
 & ($k=0.17$) & ($k=0.62$) & ($k=1.0$) \\
\hline
10  & 6.9 & 4.6 & 3.6 \\
20  & 7.5 & 1.7 & 1.2 \\
50  & 3.7 & 0.045 & 0.039\\
100  & 1.4 & 0.02 & 0.076\\
200  & 0.41 & 0.18 & 0.0079\\
300  & 0.10 & 0.12 & 0.0026\\
\hline
\end{tabular}
\end{center}
\caption{Absolute error $\Delta$ in the phase shift for different sizes of the oscillator
basis ($b=0.3$).}
\label{tab:error1d}
\end{table}

We observe that the error of the hybrid method has rather intricate and
oscillatory behavior in terms of the size, but
in general decreases with increasing oscillator basis. Full
convergence is however not yet obtained. In any case it is seen
that the approach provides relatively stable and accurate results
for different values of  the  angular
momenta, energy ranges, and sizes of the oscillator basis.

\subsection{$s$-wave benchmark with product two-body potentials}
As described above, the hybrid model introduced in this paper
is easily extended to systems with more degrees of freedom, in
contrast to the original JM method.  We demonstrate this on a model
breakup problem with a short-range potential taken from
\cite{Resgino1999pra}:
\begin{multline}
\left( -
\frac12 \frac{\partial^2}{\partial x^2} - \frac12
\frac{\partial^2}{\partial y^2} - V_0 \mathrm{e}^{-x} - V_0 \mathrm{e}^{-y} \right. \\ \left. +
W_0 \mathrm{e}^{-x-y} - E \right) \Psi = 0 
\label{eq:2d_1}
\end{multline}
with $V_0=3$ and $W_0=10$. These interaction parameters yield one
bound state for the attractive ``one-particle'' potential $V_0$ with
an energy $E_0=-0.411$.  We choose the energy of the incident particle
to be $E_{inc}=0.882$ (similarly to \cite{Resgino1999pra}) what leads
to a total energy of the system $E=0.471$.

\begin{figure}[ht]
\includegraphics[width=0.9\linewidth]{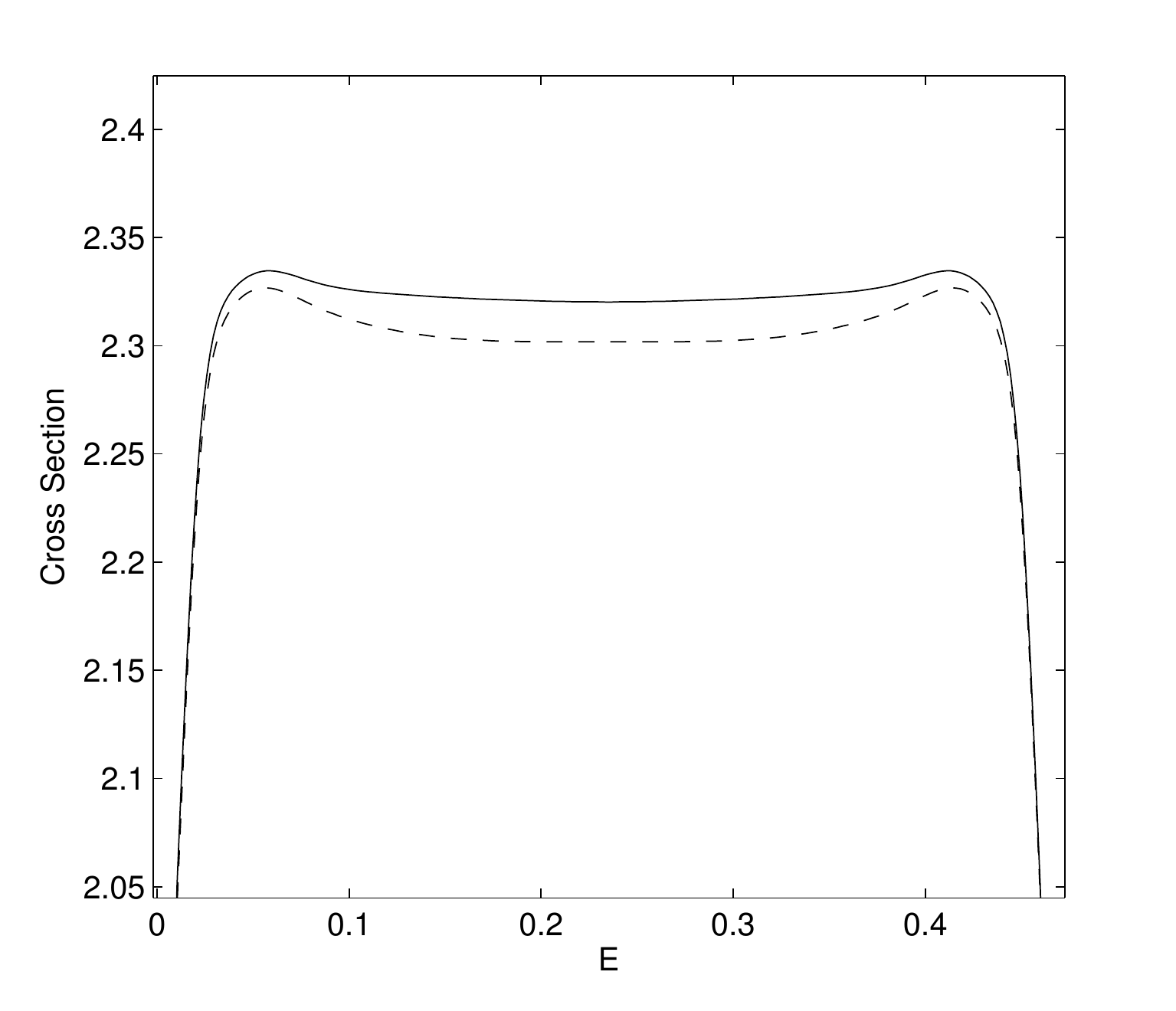}
\caption{SCDS results of $s$-wave scattering
  for model problem (\ref{eq:2d_1}), for a total energy of 0.471 (solid line --- finite differences, dashed line --- JM-ECS). The parameters of the
  calculation are $N$=80 and $b$=0.3. The
  vertical axis has been rescaled to emphasize the difference. \label{sdcs_1}}
\end{figure}

Fig. \ref{sdcs_1} pictures the single differential cross section (SDCS)
for the breakup after impact. The escaping particles have an energy
of $0.471$ to share between them.   Since the two particles are
indistinguishable the cross section is symmetric around $0.471/2$. 
In the figure we also compare with the results of the finite
difference calculations. We see a slight difference for equal energy
sharing. To highlight the difference we have scaled the vertical axis.

The error between the finite difference calculations and the hybrid
method is given in table \ref{tab:error} where we look at the error
for a particular energy sharing.  As the number of oscillator states increases the results converge.

In Table \ref{tab:error} we display the difference between the
results of the hybrid method and those of a full finite-difference calculation
in terms of the the size of the oscillator basis $N$.

\begin{table}[ht]
\begin{center}
\begin{tabular}{c|c}
\hline
$N$ & $|\Delta|$\\
\hline
10  & 0.0937\\
20  & 0.0836\\
40  & 0.0678\\
60  & 0.0592\\
80  & 0.0184\\
\hline
\end{tabular}
\end{center}
\caption{Absolute error $\Delta$ of the hybrid method w.r.t.\ a full finite-difference approach
for the SDCS results of problem (\ref{eq:2d_1}). The
  total energy is 0.471, and there is equal energy sharing between the particles.} \label{tab:error} 
\end{table}

\subsection{More realistic $s$-wave benchmark with Gaussian potential}

\begin{figure}[ht]
\includegraphics[width=0.9\linewidth]{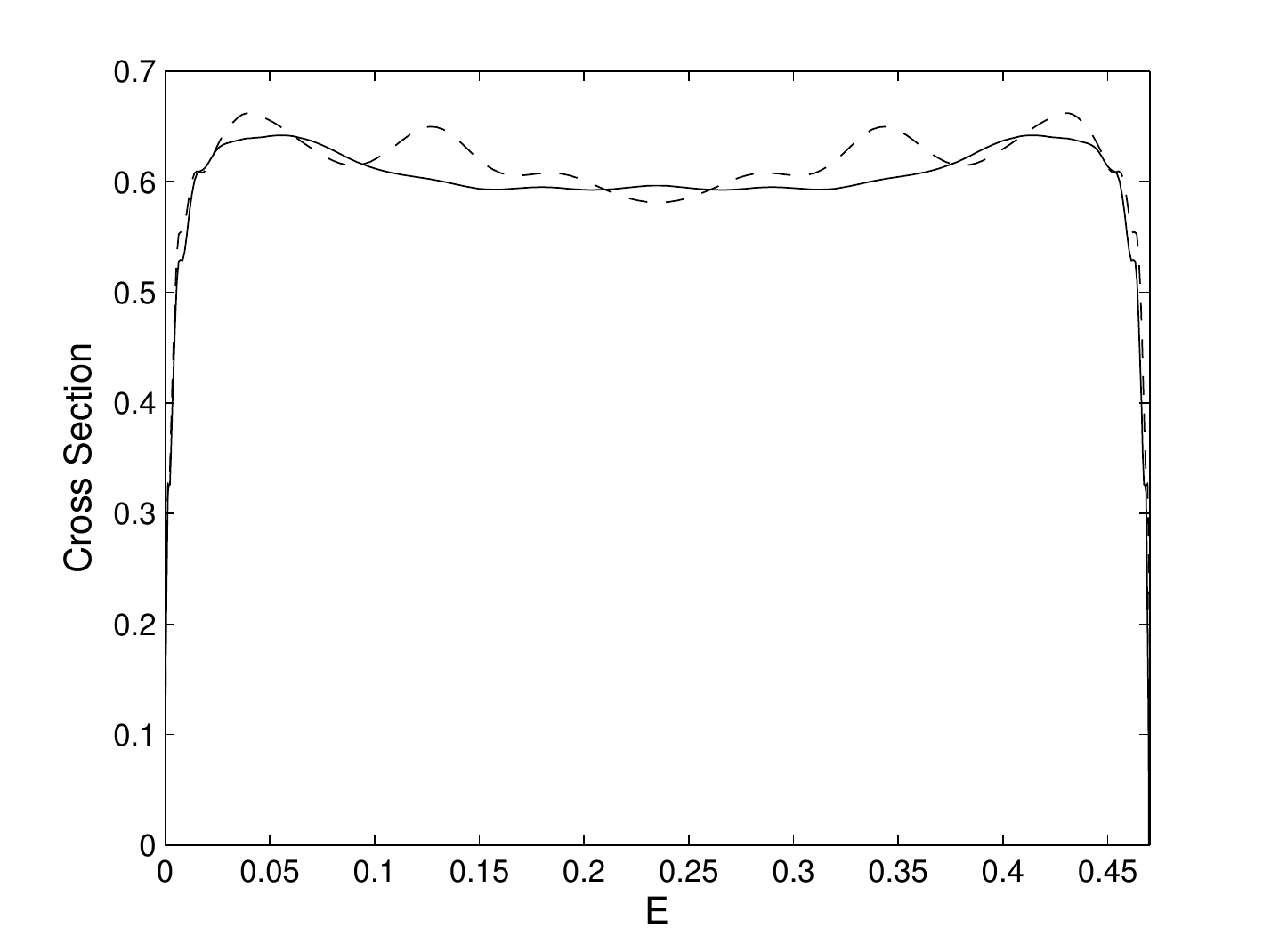}
\caption{SCDS results of s-wave scattering
  for model problem (\ref{eq:2d_2}) with a Gaussian two body potential (solid line --- finite differences, dashed line --- JM-ECS).
  The parameters of the
  calculation are $N$=70 and $b$=1.3.
  \label{fig:scds_2}}
\end{figure}

All potentials in (\ref{eq:2d_1}) are exponentially
decaying in both coordinates, and such a model is not very realistic.
In real problems the inter-particle
potential depending on the distance between particles should decay significantly
slower. We consider a more realistic three-dimensional problem, but similar
to the one described above. This could correspond to a scattering problem with two
equal, light, particles and a third heavy one:
\begin{multline}
\left( -
\frac12 \Delta_{r_1} - \frac12
\Delta_{r_2} - V_0 \mathrm{e}^{-|\mathbf r_1|} - V_0 \mathrm{e}^{-|\mathbf r_2|} \right. \\ \left. + 
W_0 \mathrm{e}^{-|\mathbf r_1 - \mathbf r_2|^2} - E \right) \Psi = 0 
\end{multline}
with $\mathbf r_1$ and $\mathbf r_2$ the coordinates of the two
(light) particles with respect to the third one. Here we consider a Gaussian
form of the potential for the interaction between the light particles to simplify the
subsequent calculations, and to obtain a faster-decaying $s$-wave potential.
The $s$-wave projection of this equation yields
\begin{multline}
\left( -
\frac12 \frac{\partial^2}{\partial x^2} - \frac12
\frac{\partial^2}{\partial y^2} - V_0 \mathrm{e}^{-x} - V_0 \mathrm{e}^{-y} \right. \\ \left. + 
W_0 \frac{\left(\mathrm{e}^{-(x-y)^2}-\mathrm{e}^{-(x+y)^2}\right)}{2xy} - E \right) \Psi = 0 
\label{eq:2d_2}
\end{multline}
In Fig.\ \ref{fig:scds_2} we show the SCDS for this problem, and compare with the results from a full finite
difference calculation. In these calculations we had to increase the size of the finite-difference grid to cover the interaction region, which leads to an increase increasing of the oscillator parameter (see \ref{eq:gridsize}).
We considered $b=1.3$ in this calculation.

Overall the results are comparable. The small
oscillations in the finite difference results come from
small numerical reflections at the point of complex scaling. These can
be eliminated by using a smooth complex scaling or a PML \cite{berenger1994perfectly} as an
absorbing boundary.  In a similar way, the hybrid method result
also shows comparable oscillations.

Table  \ref{table:errors_sdcs} shows the errors in the SDCS as a function of the oscillator basis size. Comparing the error values with the value of SDCS we can see that the relative error is considerably bigger than in previous problem. Apart of the increased complexity of the problem, the reason for this is that the asymptotic relation (\ref{eq:as_relation}) becomes less accurate with an increase of the oscillator parameter $b$.

\begin{table}[ht]
\begin{center}
\begin{tabular}{c|c}
\hline
$N$ & $|\Delta|$\\
\hline
10  & 0.3730 \\
20  & 0.0221 \\
30  & 0.0776 \\
50  & 0.0685 \\
70  & 0.0153 \\
\hline
\end{tabular}
\end{center}
\caption{
Absolute error $\Delta$ of the hybrid method w.r.t.\ a full finite-difference approach
for the SDCS results of problem (\ref{eq:2d_2}). The
  total energy is 0.471, and there is equal energy sharing between the particles.
  \label{table:errors_sdcs}}
\end{table}

\section{Application to nuclear $p$-shell scattering}

In this section we go beyond model problems and apply our method in the
more realistic context of $\alpha$-scattering in $p$-shell nuclei \cite{monquaclu}, in particular
$^8Be$ \cite{Be-paper}. 

This is a commonly used benchmark problem in nuclear cluster physics.
We will perform an $\alpha-\alpha$ scattering calculation including fully
microscopic states for the interaction region, and using a semi-realistic
nucleon-nucleon interaction. We compare the results with those obtained
in the JM approach.

In the asymptotic region the system decays into two point particles, each
corresponding to an $\alpha$-particle. At closer distances, however, the internal
structure of the clusters becomes of importance because of the microscopic
interactions and the Pauli exchanges between nucleons. 




The cluster basis states have the form
\begin{equation}
 \Psi_{nL}=\widehat{\mathcal{A}}\left\{ \Phi_{1}\left( \alpha\right)
  \Phi_{2}\left(\alpha\right) \phi_{nL}\left( \mathbf{r}\right) \right\},
\end{equation}
where $\widehat{\mathcal{A}}$ is the antisymmetrization operator, and $\Phi_{1,2}\left(  \alpha\right)$ is
a translation invariant shell-model state built up of \mbox{$s$-orbitals} for the
$\alpha$-particles. The state $\phi_{nL}\left(\mathbf{r}\right)$
is a three-dimensional harmonic oscillator state for the relative motion of
the $\alpha$-clusters
\begin{eqnarray}
  \phi_{nL}\left( \mathbf{r}\right)
   & = &\phi_{n+n_{0},LM}\left( \mathbf{r}\right)\\
   & = &\mathcal{N}_{n+n_0,L}~\rho^{L}e^{-\rho^{2}/2}L_{n+n_0}^{L+1/2}\left( \rho^{2}\right)
      Y_{LM}\left(  \widehat{\mathbf{r}}\right) \nonumber\\
  \rho & =&\frac{\left\vert \mathbf{r}\right\vert }{b},
  \quad\mathcal{N}_{nL}=\sqrt{\frac{2\Gamma\left(n+1\right)}{\Gamma\left(n+L+3/2\right)}}
  \nonumber
\end{eqnarray}
We use $n_{0}$ to denote the minimal value of the shell number allowed by the Pauli
exclusion principle
\begin{equation}
  \label{eq:nmin}
n_{0}=
\begin{cases}
(N_{min}-L)/2 & \text{for }L<N_{min},\\
0 & \text{for }L\geq N_{min},
\end{cases}
\end{equation}
where $N_{\min}$ is the number of oscillator quanta in that shell.
\begin{equation}
N_{\min}=
\begin{cases}
  A-4, &\text{normal parity states: } \pi=\left(-1\right)^{A}\\
  A-3, &\text{abnormal parity states: }\pi=\left(-1\right)^{A+1}
\end{cases}
\label{eq:nmin2}
\end{equation}
In this form $n=0,1,...$ numerates all the Pauli allowed states for the cluster
relative motion. For all further details, we refer to \cite{monquaclu} and \cite{Be-paper}.

A common nucleon-nucleon interaction used for this sytem is a Volkov N1 (V1) potential,
which can be written in the form 
\begin{equation}
  V_{ij}=
    {\displaystyle\sum\limits_{k=1}^{2}}
    V_{k}(1+mP_{ij}^{r})
    \exp\{-(r_{ij}/a_{k})^{2}\},
  \label{eq:forces}
\end{equation} 
where $m$ is the Majorana exchange parameter.
The strength parameters $V_{k}$ determine the
repulsive short-range core and the long-range
attraction. We take the same parameters as in \cite{Be-paper} and \cite{monquaclu}, but omit
the Coulomb interaction between protons to simplify the effective asymptotic
$\alpha-\alpha$ interaction.

The calculations of matrix elements in the internal region is
identical as in the JM approach. The difference lies in the asymptotic
region, where the explicit asymptotic potential, expanded in the oscillator
basis, is now replaced by the finite differences approach, as
discussed in section \ref{sec:jmecs}, after proper matching of both
regions; the asymptotic $\alpha-\alpha$ interaction reduces to the
point-like effective free particle kinetic energy between the two
clusters.

\begin{figure}[ht]
\includegraphics[width=0.9\linewidth]{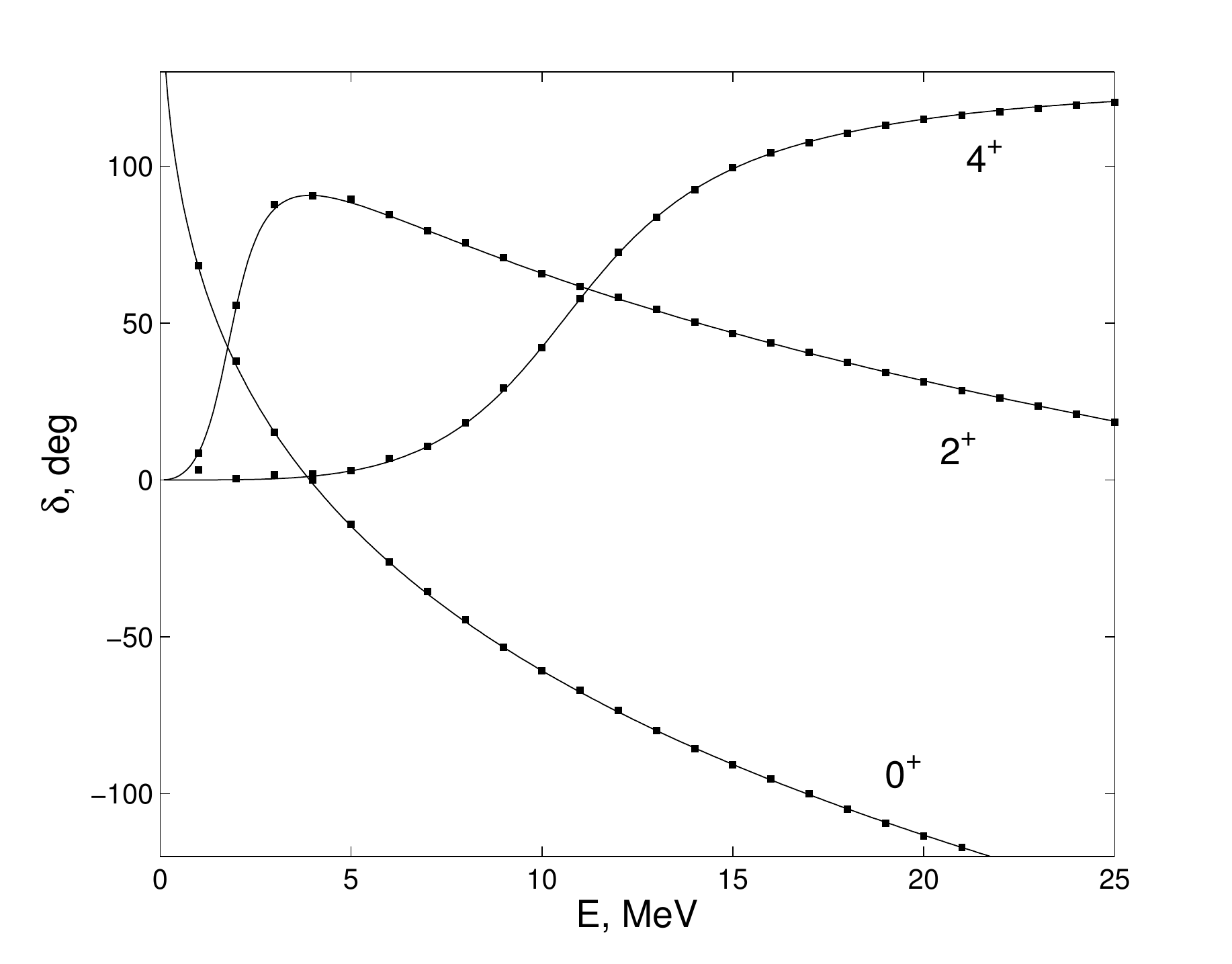}
\caption{ Scattering phase shifts for the $\alpha-\alpha$ system for
  different values of total angular momentum calculated with MJM
  (solid lines) and hybrid JM-ECS method (closed squares). All calculations
  were made with 80 oscillator states}
\label{fig:alpha}
\end{figure}

\begin{figure}[ht]
\includegraphics[width=0.9\linewidth]{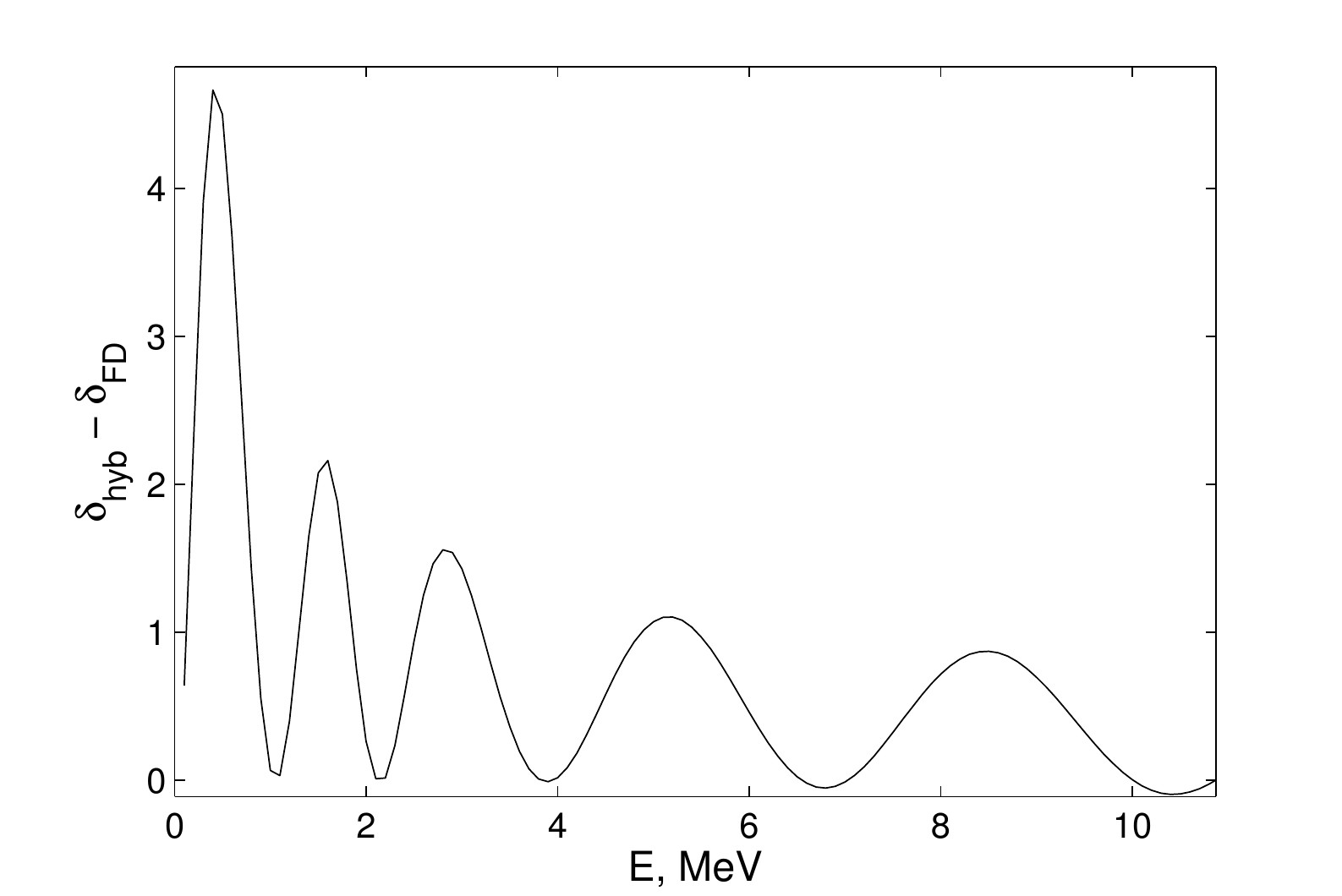}
\caption{ Difference between MJM and JM-ECS scattering phases for the 2$^+$ state of the $\alpha-\alpha$ system.}
\label{fig:phase_diff}
\end{figure}

In Fig.~\ref{fig:alpha} we show the $J^{\pi}=0^+,2^+$ and $4^+$ phase
shifts for the $^8$Be system calculated with the hybrid JM-ECS  and JM approaches.
For the latter we take results as they have been obtained in \cite{Be-paper}, where a
modified JM approach was considered, providing faster convergence  \cite{Vanroose2001}; we refer to
such calculation as MJM for consistency.
The overall results are very close to each other. The only noticeable discrepancies
occur at very low energies. The difference between MJM and JM-ECS phase
shifts in this region is presented in Fig.~\ref{fig:phase_diff}. The oscillations in
this difference correlate with the value of the derivative of the wave function in the
matching point between the two representations. The source of the error is a
``reflection'' from this point, and an improvement of the method to reduce this discrepancy is currently under research.



\section{Discussion and Conclusion}
This article reports on initial efforts to introduce absorbing
boundary conditions in quantum scattering problems where the wave function is 
represented by a finite sum of oscillator states.  We have
introduced the exterior complex scaling (ECS) boundary conditions by
constructing a representation of the wave function that combines the
oscillator states with a grid based representation.  The oscillator
representation covers the inner region of the problem, where the main
interaction occurs, while the grid covers the near field.  The two
regions are numerically matched at an interface using an asymptotic
expression for the expansion of oscillator states.

The far field amplitude, which gives the cross section, is extracted
from the numerical wave function using a surface integral. Extra care
is required in the calculation of this integral in this
representation.

We have numerically solved several benchmark problems and compared our
results with literature and plain ECS calculations. 

Our results agree with existing methods and confirm that the proposed
method works. However, the accuracy still needs to be improved.  This
can be done by including mixed potential matrix elements from
different representations, by using a different finite difference grid,
or by using a higher order matching condition at the interface between
the oscillator representation and the finite difference
representation. Also a proper treatment of the Coulomb interaction and other effective interactions with long-range behavior must be considered in the future development of the method.

The building blocks presented in this papers are the starting point
for a fully coupled calculations similar to
\cite{vanroose2005complete}. In such a calculation the wave function
is a sum over $l_1$, $m_1$ and $l_2$, $m_2$ of 2D functions
$\psi_{l_1m_1,l_2m_2}$ combined with $Y_{l_1m_1}(\Omega_1)
Y_{l_2m_2}(\Omega_2)$. The linear system is then a coupled system
where each diagonal block is a system like
eq. \eqref{eq:twodimensional}.  From such a representation of the wave function we can extract not
only the single differential cross section, but also the triple
differential cross section, which gives the amplitude for certain
breakup directions.  In a similar way, the method can be applied to
different multi-channel calculations of three-cluster systems and different reaction processes with light nuclei. In all such problems asymptotic description can be greatly simplified if we use absorbing boundary conditions.

We conclude that it is possible to introduce a boundary layer that
absorbs all outgoing waves within a spectral basis such as the oscillator representation.

\begin{acknowledgments}
This work is supported by FWO Flanders G.0.120.08.
\end{acknowledgments}

\end{document}